\documentclass[aps,twocolumn,pra,tightenlines,floatfix,showpacs]{revtex4}
\usepackage[dvips]{graphicx}
\usepackage[english]{babel}
\usepackage{amsmath}
\usepackage{amssymb}
\usepackage{times}
\usepackage{bm,times}
\newcommand{\bs}{\begin{split}}
\newcommand{\es}{\end{split}}
\usepackage[dvips]{graphicx}

\newcommand{\mb}[1]{{\mathbf{#1}}}
\newcommand{\be}{\begin{equation}}
\newcommand{\ee}{\end{equation}}
\newcommand{\ba}{\begin{eqnarray}}
\newcommand{\ea}{\end{eqnarray}}
\newcommand{\ek}{\epsilon_{\mathbf{k}}}
\newcommand{\Ek}{E_{\mathbf{k}}}
\newcommand{\uk}{u_{\mathbf{k}}}
\newcommand{\vk}{v_{\mathbf{k}}}
\newcommand{\xik}{\xi_{\mathbf{k}}}
\newcommand{\sumk}{\sum_{\mathbf{k}}}

\newcommand{\sumq}{\sum_{\mathbf{q}}}

\newcommand{\p}{\partial}

\newcommand{\vvk}{\mathbf{k}}
\newcommand{\vq}{\mathbf{q}}
\newcommand{\Ekk}{E_{\mathbf{k}}^{o}}

\def\ket#1{|#1\rangle}

\newcommand{\vect}[1] {\mathbf{#1}}

\newcommand{\up} {\uparrow}
\newcommand{\down} {\downarrow}

\begin{document}

\title{Comparison of Different Pairing Fluctuation Approaches to
BCS-BEC Crossover}

\author{$^{1}$K. Levin, $^{1,2}$Qijin Chen, $^{1}$Chih-Chun Chien and
$^{1}$Yan He }

\affiliation{$^1$James Franck Institute and Department of Physics,
University of Chicago, Chicago, Illinois 60637, USA}
\affiliation{$^2$Zhejiang Institute of Modern Physics and Department of
Physics, Zhejiang University, Hangzhou, Zhejiang 310027, China}

\date{\today}

\begin{abstract}

The subject of BCS - Bose Einstein condensation (BEC) crossover is
particularly exciting because of its realization in ultracold Fermi gases
and its possible relevance to high temperature superconductors. In
the paper we review that body of theoretical work on this subject
which represents a natural extension of
the seminal papers by Leggett and by
Nozi\`eres and Schmitt-Rink (NSR).
The former addressed only the ground state, now known as the
``BCS-Leggett" wave-function and the key contributions of the
latter pertain to calculations of the superfluid transition temperature $T_c$.
These two papers have given rise to two main and, importantly,
distinct, theoretical schools in
the BCS-BEC crossover literature.
The first of these extends the BCS-Leggett ground state
to finite temperature and the
second extends
the NSR scheme away from
$T_c$ both in the superfluid and normal
phases.
It is now rather widely accepted that these extensions of NSR
produce a different ground state than that first introduced by Leggett.
This observation provides a central motivation for the
present paper which seeks to clarify the distinctions in the two
approaches.
Our analysis shows how 
the NSR-based approach views the bosonic contributions more completely
but it treats the fermions as ``quasi-free''.
By contrast, the BCS-Leggett
based approach treats the fermionic contributions more completely 
but it treats the bosons as ``quasi-free''.
In a related fashion, the NSR based schemes approach the
crossover between BCS and BEC by starting from the BEC limit and the
BCS-Leggett based scheme approaches this crossover by starting from
the BCS limit.
Ultimately, one would like to combine these two schemes.  There are,
however, many difficult problems to surmount in any attempt to bridge
the gap in the two theory classes.  In this paper we review the
strengths and weaknesses of both approaches.  The flexibility of the
BCS-Leggett phase and its ease of handling make it more widely used in
$T=0$ applications, although, the NSR-based schemes are more widely
used at $T \neq 0$.  To reach a full understanding, it is important in
the future to invest effort in investigating in more detail the
$T=0$ aspects of NSR-based theory and at the same time the $T \neq 0$
aspects of BCS-Leggett theory.

\end{abstract}


\maketitle

\section{Introduction}
\label{sec:1}

The subject of BCS-Bose Einstein condensation (BEC) crossover has
recently become an extremely active research area. This is due
principally to the discovery \cite{Jin3,Grimm,Jin4,Ketterle3,KetterleV,
  Thomas2,Grimm3,ThermoScience,Salomon3,Hulet4} of superfluid phases in
ultracold Fermi gases which exhibit this
crossover.  Adding to the importance of this work is the
view espoused by a number of theorists
\cite{randeriareview,ourreview,Varenna,LeggettNature,Strinaticuprates} that the
high temperature superconductors are mid-way between BCS and BEC.  Now,
with an unambiguous realization of this scenario in the fermionic
superfluids, one has the opportunity to investigate this physical
picture more closely and, it is hoped, gain insight into the cuprate
superconductors.  Equally exciting is the opportunity to generalize, 
and in the process, gain insight into what
is arguably the paradigm for all theories in condensed matter physics:
Bardeen Cooper Schrieffer theory.  For all these reasons a large number
of variants of BCS-BEC crossover theory have been suggested in the
literature.  It is the purpose of the present paper to present an
overview of two main classes of theories, discussing their strengths and
weaknesses. Contrasting and comparing different approaches will,
hopefully, point to new directions for future theoretical and
experimental research.

Initial theoretical work \cite{Eagles,Leggett} on the subject of BCS-BEC
crossover focussed on a ground state which was shown to be the same as
that proposed by Bardeen Cooper and Schrieffer, when it is extended
to accomodate a
continuous evolution from BCS to BEC. We call this the ``BCS-Leggett"
state. Here the chemical
potential is solved self consistently as the attractive interaction
strength is varied. In this way it became clear that the BCS trial
wavefunction was far more general than was originally thought.  Somewhat
later, Nozi\'eres and Schmitt-Rink (NSR) \cite{NSR} presented a scheme
for calculating the transition temperatures $T_c$, which made
the case that the evolution from BCS to BEC was again continuous at
finite temperature.  

The discovery of high temperature superconductivity
and the observation that their coherence length $\xi$ (or pair size) was
anomalously small led T. D. Lee and R. Friedberg to argue that one
should include bosonic degrees of freedom in addressing high $T_c$
superconductors.  These authors introduced \cite{TDLee1,TDLee2} the
``boson-fermion'' model almost immediately after the discovery of
cuprate superconductivity.  In a similar vein, Randeria and co-workers
\cite{randeriareview} proposed that the NSR scheme might be directly
applicable to these exciting new materials.  Subsequently other
theorists have applied this BCS-BEC crossover scenario to the high $T_c$
cuprates \cite{Chen2,Micnas1,Ranninger,Strinaticuprates}.  Additional
support has come from the experimental condensed matter community among
whom a number \cite{Uemura,Renner,Deutscher,Junod} have presented data
which can be interpreted within this picture.  Adding to the enthusiasm
is the observation of a ubiquitous (albeit controversial) ``pseudogap''
phase \cite{LeeReview,ourreview,Varenna} in the underdoped cuprates, which was
argued \cite{RanderiaVarenna,Janko} to be consistent with a BCS-BEC
crossover scenario.

The characterization of pseudogap effects associated with BCS-BEC
crossover was, in fact, a crucial step. It was first recognized that one
should distinguish the pair formation temperature $T^*$ from the
condensation temperature $T_c$ \cite{MicnasRMP,randeriareview}.  That
the magnetic properties of the normal phase in the temperature regime
between $T_c$ and $T^*$ would be anomalous was pointed out on the basis
of numerical calculations, on a two dimensional lattice. Here it was
found that the spin susceptibility was depressed at low temperatures
\cite{Trivedi} and this depression was associated with a ``spin gap''
which is to be distinguished \cite{RanderiaVarenna}
from a pseudogap which affects the ``charge channel" as well.
The fact that BCS-BEC crossover theory was, indeed, associated with a
more general form of
pseudogap, thus, required further analysis and calculations.  Using the
formalism of the present paper, subsequent, theoretical studies of the
spectral function (both above \cite{Janko} and below \cite{Chen4} $T_c$)
and the superfluid density \cite{ChenPRL98} showed that
a normal state pairing gap appeared in
\textit{both} the spin and charge channels and, furthermore, affected
the behavior below $T_c$ as well \cite{ChenPRL98} as above.

The BCS-BEC crossover approach was applied to the ultracold Fermi gases, by Holland and
co-workers \cite{Milstein} and by Griffin and Ohashi \cite{Griffin} in
advance of the discovery of fermionic superfluidity. 
The two groups predicted that
the magnetic field tuneability associated with an atomic Feshbach resonance
would lead to an unambiguous realization of the crossover scenario.
These earliest applications to ultracold Fermi gases considered a
Hamiltonian rather similar to the ``boson-fermion'' model of Lee and
co-workers \cite{TDLee1,TDLee2} where the bosons were related to 
the so-called closed channel molecules
of the Feshbach resonance and the fermions
with the open channel.  Subsequent work has shown that these
two channel complications can be essentially ignored so that the description of
Fermi gas superfluidity is addressed using the same, simpler (or one
channel) model as was used in the cuprates.

There is now a fairly extensive theoretical literature
\cite{ourreview,Varenna,Griffin2,Strinati2} on the Fermi gas
superfluids. Nevertheless, there are two main theoretical schools, which
have emerged. These
address a wide variety of different issues and experiments.  The
first of these builds more directly on the BCS-Leggett ground state and
its finite temperature extensions
\cite{Janko,Kosztin1,Chen1,JS2,Torma2}.  The second approach
\cite{Griffin,Milstein,Strinati,Strinati2,Griffin2,Strinati3,Strinati4}
builds on the contribution of Nozieres and Schmitt Rink which addressed a
calculation of the transition temperature $T_c$. The NSR scheme has been
extended by these and other authors away from $T_c$ both in the
superfluid and normal phases. \textit{It is now rather widely accepted
  that these extensions of NSR produce a different ground state
  \cite{ourreview,Strinati5,Randeriaab} than that first introduced by
  Leggett \cite{Leggett} and by Eagles \cite{Eagles}}.  This observation
provides a central motivation for the present paper.  We want to set
down our current understanding of what is known about the NSR-based
theories from zero to very high $T$ and similarly, address how the
simplest ground state of BCS-Leggett evolves with increasing temperature
away from zero. Since the ground states are different, we can safely
assume that the finite $T$ behavior is as well.  It should be stressed
that while we attribute these author group names to the two different
schools, the eponymous authors are \textit{not} the origin of the theory
reviewed here. The original paper by Leggett was only concerned with the
ground state and that by Nozieres and Schmitt-Rink recapitulated and
expanded on the results of Leggett and then went on to compute $T_c$
using an approach which was not associated with this same (BCS-Leggett) $T=0$
state.

The task in any finite temperature crossover theory is to arrive at a
characterization of the thermal excitations of both the normal and
superfluid phases.  From this analysis all transport and thermal
properties can in principle be obtained. Without any detailed
microscopic theory one can still anticipate the general features of
BCS-BEC crossover theory. In the BCS regime and below $T_c$, the
excitations are the usual fermionic quasi-particles with an excitation
gap equivalent to the order parameter. This gap represents the energy cost
of unbinding the condensate pairs. By contrast, above $T_c$ this gap is
absent and the normal state is a Fermi liquid. In the BEC regime, it is
energetically unfavorable to break up the pairs and so the excitations
are purely bosonic above and below $T_c$. In the superfluid phase, they
are, moreover, gapless.  In between, in the interesting unitary regime, 
the excitations are expected to be a mix of fermionic and bosonic
character. Here, importantly even the normal state has some bosonic
features associated with the formation of ``pre-formed pairs''.  These
pairs arise from stronger than BCS attractive interactions.  As a
consequence there is an excitation (pseudo)gap for fermionic excitations 
which appears above $T_c$.  With
progressively lower temperatures below $T_c$, more and more of these
pairs drop into the condensate.  The challenge then is to treat the
strongly interconnected bosonic and fermionic degrees of freedom in the
most physically correct fashion.

Below $T_c$ the two schools referred to above emphasize different
aspects of this picture.  One can summarize in the simplest fashion the
key differences.  The NSR-based approach views the bosonic contributions
more completely but it treats the fermions as ``quasi-free''.  Many body
effects are effectively absent in the fermionic dispersion relation
(called $\Ek^o$, 
which appears in the counterpart gap equation), except
via a renormalization of the fermionic chemical potential $\mu^*$.  The
BCS-Leggett based approach treats the fermionic contributions more
completely than the alternative approach but it treats the bosons as
``quasi-free''. Many body effects in the bosonic dispersion, (which we
call $\Omega_{\bf q }^o$), are absent in the gap equation except via an
effective pair mass renormalization $M^*$.

More specifically, the NSR approach incorporates a linear dispersion in
the bosonic degrees of freedom at small wavevector which is associated
with the collective mode spectrum of the condensate.  However, $T_c$ is calculated in the
same way as for non-interacting fermions, except for the renormalization
in $\mu^*$.  The BCS-Leggett based approach 
in effect approximates the bosonic
degrees of freedom associated with the non-condensed pairs. While the
collective modes of the order parameter have a linear dispersion, the
non-condensed pairs have a quadratic dispersion and represent otherwise
free ``bosons''.  Here, $T_c$ is calculated in the
presence of a pseudogap so that the condensing fermionic quasiparticles
have admixed bosonic character.

It is reasonable to conclude that the NSR based scheme approaches the
crossover between BCS and BEC by starting from the BEC limit and the
BCS-Leggett based scheme approaches this crossover by starting from the
BCS limit.  For the former, indeed, the boson-like propagators which one
deduces are found to have many similarities to Bogoliubov theory for
true bosons.
It is claimed \cite{Strinati6} that the NSR-based approach is most
accurate at temperatures low compared to $T_c$, presumably because there
the bosonic degrees of freedom are those associated with the condensate
and its collective modes.  Thus, it is reasonable to assume that this
represents the better ground state.  By contrast the BCS-Leggett based
scheme is more suitable at moderate temperatures within the superfluid
phase and up to the pairing onset temperature $T^*$. Above $T^*$, the
two approaches can be viewed as equivalent.

Ultimately, one would like to combine these two schemes.  There are,
however, many difficult problems to surmount in any attempt to bridge
the gap in the two theory classes. Not only is it difficult to effect
such a combination but, thus far, there is no mean field theory of a weakly
interacting Bose gas \cite{Griffin4,AndersenRMP} which addresses the
entire regime from $T=0$ through and above $T_c$ and which does not have
at the same time a problematic first order transition.  Thus, for the
interacting Bose gases, there is no counterpart of BCS theory which
works so well over the entire temperature range.  These complications,
in true Bose systems, appear to be transmitted to NSR based theories of
the Fermi gases.
These spurious first order
transitions \cite{firstordertransitionpapers} can lead to derivative
discontinuities in the density profiles at the condensate edge and
non-monotonic or discontinuous behavior in the superfluid density, even
in the intermediate or unitary regime.  BCS theory, by contrast,
exhibits none of these effects.
If one is to find a smooth crossover between BCS and
BEC at all temperatures $T$ these issues will need to be overcome.

Additional problems appear if one tries to bridge the gap by starting
with the BCS-Leggett based scheme. The first task is to establish how
non-condensed pair effects modify the collective mode
spectrum.  This appears to be a difficult problem. While, some progress
has been made \cite{Kosztin2} towards computing pseudogap effects on the
Anderson Bogoliubov mode, there is, however, an even greater difficulty
in coupling the non-condensed pairs with the renormalized collective
modes.  To arrive at this hybridization, one needs to introduce
boson-boson coupling which requires that one go beyond the simple
T-matrix scheme which one considers in addressing the non-condensed
component.  This is not to say that the coupling between condensed and
non-condensed pairs is absent, it must be there in the ultimate
theory, but it will be difficult to implement.

To summarize, the ground state produced by NSR-based approaches is
likely to represent an improvement over that of the BCS-Leggett based
approach, particularly when it comes to quantitative comparisons, and
most particularly when the system is on the BEC side of resonance.
However at the semi-quantitative or qualitative level one is often
required to consider the BCS-Leggett ground state, and its
finite temperature extensions, since globally this state behaves more
smoothly. Moreover, this state is easier to handle and can accomodate
inhomogeneities via Bogoliubov deGennes theory
\cite{Machida,Chienvortex,Hovortex,BdG_Imbalance}. It is the also the
primary way to study phases with population imbalance
\cite{SR06,SM06,Chien06,Rice2,ChienPRL}, particularly in the presence of
a trap.

An understanding of BCS-BEC crossover provides an excellent vehicle for
reviewing the central features of two types of mean field theories:
strict BCS theory and the theory(s) of the weakly interacting Bose gas.
In both systems there is the potential for carrying some confusion over
to the crossover problem, since there are important ``degeneracies''
which are not general and which occur at each endpoint. In strict BCS
theory the order parameter $\Delta_{sc}$ is the same as the excitation
gap $\Delta$. This relationship cannot persist in BCS-BEC crossover.
In the Bogoliubov
theory of the weakly interacting Bose gas the collective mode frequency
is the same as the single particle excitation energy.  This degeneracy
derives from the coupling between the
order parameter and single particle excitation spectrum. This situation
is not the case in BCS theory. The way in which the linearly
dispersing order parameter collective modes interact with the
quasi-particle (fermionic) excitations and the extent to which they
couple is subtle in BCS theory.

Indeed, if one applies the Landau criterion to a magnetically
dirty, gapless superconductor, (where importantly it is found that
$\Delta \neq \Delta_{sc}$) it must, of course, reveal that
superconductivity is stable.  This Landau criterion should
\textbf{not}, then, refer to all possible excitations of the system but only
those which couple to the condensate, that is, associated with the
density fluctuations \cite{GuyDeutscher}.  The gapless single particle
excitations do not compromise superfluidity and thus one can presume that
they do not couple directly to the collective modes. An analogous
inference can, then, be made about a clean BCS superconductor which
suggests a decoupling between the condensed and non-condensed
components-- at the strict BCS level.

There is another avenue for confusion.  The flexibility of the
BCS-Leggett phase and its ease of handling
make it more widely used in $T=0$ applications, although, the NSR-based
schemes are more widely used at $T \neq 0$.  One has seen just this
dichotomy in the original paper \cite{NSR} by Nozieres and Schmitt-Rink.
\textit{To reach a full understanding, it is important, then, to invest
  some effort in investigating in more detail the $T=0$ aspects of
  NSR-based theory and at the same time the $T \neq 0$ aspects of
  BCS-Leggett-based theory}.

The remainder of the paper is divided into four sections.
Section \ref{sec:2} presents a theoretical overview of
BCS-BEC crossover theory beginning first with an alternative
presentation of strict BCS theory at general temperatures $T$,
which provides general insights.
Then a brief overview of the ground state equations for
the BCS-Leggett approach is presented. 
Sections \ref{sec:3} and
\ref{sec:4} give a more detailed description of the BCS-Leggett
and Nozieres, Schmitt-Rink theretical schools, respectively, at general
temperatures $T$. There we review the general equations and the specific
application to the BEC limit as well as the superfluid density. Other
issues are discussed as well which pertain to special features of
each of the two schools. These
sections are more technical and they can be skipped by
a reader so inclined who is advised to go directly to
Section \ref{sec:5}. 
Section \ref{sec:5}
summarizes crucial comparisons between the two theoretical schools.
Many of these are presented in the form of two tables. In addition
we compare plots of the transition temperature in a homogeneous
and trapped configuration and plots of the density profiles.
Our conclusions are summarized in
Section \ref{sec:6}.

\section{Theoretical Overview}
\label{sec:2}

\subsection{Ground State Wavefunctions}
\label{sec:2a}

We begin with a summary of possible ground state wavefunctions for
describing BCS-BEC crossover. The simplest one is that of BCS-Leggett
\begin{equation}
\Psi_0=\Pi_{\bf k}(\uk+\vk 
c_{\vect
k\up}^\dagger
c_{-\vect k\down} ^\dagger
|0\rangle ,
\label{eq:1a}
\end{equation}
where $v_{\bf k}$ and $u_{\bf k}$ are variational parameters.
If we define $\alpha_{\bf k} = \vk / \uk $ we may write
\begin{equation}\label{eq:BCS2}
\Psi_0 \propto \exp\Bigl(\sum_\vect k\alpha_{\vect k}^{} 
b_{0,k} ^\dagger
\Bigr)\ket{0}.
\end{equation}
Note that this state represents an essentially ideal Bose gas treatment of
the pair degrees of freedom in the sense that it can be written entirely
in terms of a single ``Bose'' operator with net zero momentum
\begin{equation}
b_{0,k} ^\dagger \equiv  c_{\vect
k\up}^\dagger
c_{-\vect k\down} ^\dagger .
\label{eq:3}
\end{equation}

One can also contemplate something closer to a Bogoliubov-level wave
function which can be simply written for the case of point bosons. 
A reasonable ansatz is:

\begin{equation}
\ket{\psi_\text{Bogoliubov}}=\exp\biggl(b_0^\dagger b_0 +\sum_{\vect q>0}
x_\vect q b_{\vect q}^\dagger b_{-\vect q}^\dagger\biggr)\ket{0}.
\label{eq:bogol}
\end{equation}

For the fermionic system, a natural extension, which has been discussed
in the literature \cite{Shina2} can be written as

\begin{widetext}
\begin{equation}\label{eq:psi}
\ket{\psi_1}=\exp\biggl(\frac{1}{2!}
\sum_{\vect K}\alpha_{\vect K}^{}c_{\vect K}^\dagger c_{-\vect K}^\dagger
+\frac{1}{4!}\sum_{\vect K\text{'s}}
\beta_{\vect K_1 \vect K_2 \vect K_3 \vect K_4}^{}
c_{\vect K_1}^\dagger c_{\vect K_2}^\dagger c_{\vect K_3}^\dagger c_{\vect K_4}^\dagger
\biggr)\ket{0},
\end{equation}\end{widetext}
where each $\vect K_i$ represents a shorthand notation for $\vect
k_i\sigma_i$, and $-\vect K$ refers to a reversal of both the momentum
and spin.
In actuality, it has been shown that to recover a consistent treatment
of Lee-Yang contributions, and to include the exact constraint on the
inter-boson scattering length \cite{Petrov}, it is necessary to keep
terms of the form
$\frac{1}{6!}\sum_{\vect K\text{'s}} \gamma_{\vect K_1\cdots\vect
  K_6}^{} c_{\vect K_1}^\dagger\dots c_{\vect K_6}^\dagger$.

We stress that this Bogoliubov-based wavefunction is \textit{not}
the basis for extended NSR theories. Nevertheless,
this hierarchy of ground states should underline the observations made
above, that we are dealing with two different and complementary
treatments of the bosonic degrees of freedom, when we investigate these
two different approaches to BCS-BEC crossover theory.
It should be stressed that, despite some confusion
in the literature, bosonic contributions are present in
the BCS-Leggett scheme as well, but they are appear as less
strongly correlated than their counterparts in the NSR scheme.
This point is re-inforced by a discussion of the BEC limit
in 
Section \ref{sec:3a}. This point is also reinforced by a
recognition of the extensive fluctuation literature in BCS
superconductors (at low dimension), which bears strong
similarity \cite{JS} to our discussion of the BCS-Leggett approach.

\subsection{Strict BCS Theory and BCS-Leggett Ground State}
\label{sec:2b}

We begin by recasting strict BCS theory in a slightly different way
which replaces the usual Gor'kov F functions with the product of one
dressed and one bare Green's function. This alternate representation
builds a basis to extend to the BCS-Leggett phase.
We define the T-matrix for a BCS superfluid as
\begin{equation}
t_{sc} (Q) = - \Delta_{sc}^2 \delta(Q) / T ,
\label{eq:2c}
\end{equation}
where $Q$ is a four-vector and $\Delta_{sc}$ is the superfluid
order parameter. This leads to the fermionic self energy, given by 
\begin{equation}
\Sigma^{BCS}(K) = \sum_Q t_{sc}(Q) G_0(Q-K)
\label{eq:2ca}
\end{equation}
so that $\Sigma^{BCS}(K) = -\Delta_{sc}^2 G_0(-K)$. 
Here, and throughout, $G_0$ is the Green's function of the
non-interacting system.  We write
\begin{equation}
G^{BCS}(K) \equiv [ G_0^{-1}(K) - \Sigma^{BCS}(K) ]^{-1}.
\label{eq:12c}
\end{equation}
The well known BCS gap equation is:
\begin{equation}
 1+ U 
\sum_K G^{BCS}(K)G_0(-K)
= 0,~~T \leq T_c ,
\label{eq:5c}
\end{equation}
which can be written in the more familiar form 
\begin{equation}
  \Delta_{sc}(T) =-U \sum_{\bf k} \Delta_{sc}(T) \frac{1-2 f(\Ek^{BCS})}{2 \Ek^{BCS}} ,
\label{eq:8c}
\end{equation}
where $U$ is the attractive interaction which drives superfluidity. Here
\begin{equation}
\Ek^{BCS} = \sqrt{ (\ek -\mu)^2 + \Delta_{sc}^2 (T) } ,
\label{eq:9c}
\end{equation}
where $\ek = k^2 / 2m$ is the bare fermion dispersion.

Once the self energy is known, the two-body or transport
properties are highly constrained through gauge invariance or Ward
identities. For convenience, we work in the transverse gauge.  The
response kernel for a fictitious vector potential $\bf{A}$
in an isotropic system is given by
\begin{equation}
K(Q) = \frac{n}{m} - P(Q) ,
\label{eq:kernel}
\end{equation}
where $P(Q)$ is the current-current correlation function and
we have $\bf{J}(Q) = K(Q) \bf{A}$.  Now,
following the standard procedure one uses a Ward identity to construct a
consistent form for the correlation function $P^{BCS}(Q) =$
\begin{eqnarray}
&-& \frac{2 }{3 m^2} \sum_{K} \left(k + \frac {q}{2}\right)^2[G(K)G(K+Q) \nonumber \\
&+& \Delta_{sc}^2 G(K)G_0 (-K) G(K+Q) G_0 (-K-Q)],
\label{eq:response}
\end{eqnarray}
where, for convenience, we have dropped the superscript $BCS$ which must
appear on all dressed Green's functions. 

The second term in Eq.~(\ref{eq:response}) is important here. One can
represent this diagrammatically as in a ``Maki-Thompson'' diagram. More
traditionally this is written as the product of two Gor'kov F functions.
After analytic continuation ($Q\rightarrow(\Omega,{\bf q})$) and taking
$\Omega\rightarrow 0$ then $q\rightarrow 0$, this expression leads to
the usual BCS result for the superfluid density
\begin{equation}
  \frac{ n_s^{BCS}}{m} = \frac{4}{3m^2} \sum_{K}\frac {k^2\Delta_{sc}^2} {\left[\omega_{n}^2+(E^{BCS}_{\bf k})^2\right]}.
\label{eq:nsBCS}
\end{equation}

Another important collective feature of the BCS superfluid state
is the dispersion $\Omega_q = cq $ for the Goldstone Boson
which is given by solving
\begin{eqnarray}
0 &=& \frac {2}{U}
+ \sum_K [G(K)G(K+Q) \nonumber \\
&+& \Delta_{sc}^2 G(K)G_0 (-K) G(K+Q) G_0 (-K-Q)] .
\label{eq:coll}
\end{eqnarray}
Note that the four Green's functions in
Eq.~(\ref{eq:coll}) are very similar to their counterparts in
the superfluid density. This underlines the fact that the
dynamics associated with BCS theory involves
inter-pair interactions, but
only within the condensate.

We end this section
by using this analysis to write the central $T=0$ equations
for BCS-Leggett theory.
The gap equation is that of strict BCS theory at T=0 and the only
difference is that it is solved in the presence of a self consistent
equation for the fermionic chemical potential $\mu$, which must vary as
the attractive interaction $U$ varies:
\begin{equation}
\Delta_{sc}(0) =-U \sum_{\bf k} \Delta_{sc}(0) \frac{1}{2 \Ek^{BCS}}\;.
\label{eq:8f}
\end{equation}
with
\begin{equation}
n= \sum_{\bf k} \left[ 1 - \frac{\epsilon_{\bf k}-\mu} {\Ek^{BCS}}
\right].
\label{eq:12ca}
\end{equation}
Importantly, we note that an equation analogous to
Eq.~(\ref{eq:coll})
can also be used throughout the crossover
as the basis for addressing collective behavior of the order parameter such
as the superfluid density and condensate sound mode
\cite{Cote,Randeria2,Kosztin2}.
In the BCS regime this yields
$c(T=0) = v_F / \sqrt{3}$. , while in the BEC
limit $ c(T=0) \approx \sqrt {(4 \pi n a_B / M_B^2)}$.  We define the
inter-boson scattering length $a_B \equiv 2 a$ and $M_B \equiv 2m$.

All of this is relevant to the following observations.
One might be concerned that, since the BCS wavefunction seems to treat
the pairs or ``bosons'' at a cruder level than associated with the
counterpart Bogoliubov wavefunction, that this
quasi-ideal gas behavior would somehow destabilize superfluidity. This
presumption is based on the observation that an ideal Bose gas cannot be
a superfluid.  We have now seen that effects appearing in the collective
behavior associated with the condensate, such as the 
speed of sound, do not correspond to those of an ideal Bose gas.
We thus infer that the condensate can reflect a rather complex dynamics,
through the effective incorporation of higher order Green's functions
into the generalized linear response.

\subsection{Characterizing the Fermionic Degrees of Freedom in BCS-BEC  Crossover at General $T$}
\label{sec:2c}

The above summary based on strict BCS theory provides an underlying
basis for describing the \textit{fermionic} degrees of freedom in both
theoretical approaches to BCS-BEC crossover.  We emphasize that the
bosonic degrees of freedom are absent at this level and that the
fermionic degrees of freedom are not treated in an equivalent fashion in
the two theoretical schools, although some of the expressions representing
the fermions look
rather similar. 

We write for the ``gap'' and ``number'' equations

\begin{equation}
1  + U  \mathop{\sum_{\bf k}}  \frac{1 - 2 f(E_{\bf k}^{mf})}{2
E_{\bf k}^{mf}}  = 0,
\label{eq:gap_equationmf}
\end{equation}

\begin{equation}
n  = \sum _{\bf k} \left[ 1 -\frac{\ek - \mu}{E_{\bf k}^{mf}} 
+2\frac{\ek - \mu}
{E_{\bf k}^{mf}}
f(E_{\bf k}^{mf}) 
\right] ,
\label{eq:7a}
\end{equation}
where $mf$ corresponds to ``mean field'' and the fermionic dispersion is 
\begin{equation}
  \Ek^{mf} (T) \equiv \sqrt{ (\ek -\mu)^2 + \Delta_{mf} ^2 (T) } .
\label{eq:dispersionmf}
\end{equation}
This system of equations has been used by both schools to find a
reasonable estimate for the temperature at which pairing or the
pseudogap first occurs. This is called $T^*$, (which satisfies
$T_c \leq T^*$) and can be computed by
solving for the transition temperature in the strict mean field
equations.

We will show that this mean field theoretic approach with
\begin{eqnarray}
\Delta_{mf}(T) &=& \Delta(T) 
\label{eq:new1} \\
\Delta^2(T) &=& \Delta_{pg}^2 (T) + \Delta_{sc}^2(T)
\label{eq:sum}
\end{eqnarray}
is associated with the finite temperature extension of the BCS-Leggett
theory.  We have argued that the ground state wave function, Eq.~(\ref{eq:1a}), 
must necessarily also contain bosonic excitations.  
This can be seen most clearly when we examine the BEC limit
in 
Section \ref{sec:3a}.
Therefore, within this
theoretical school, one must not presume that the mean field gap is
equivalent to the order parameter.  These bosonic excitations are
accomodated by decomposing $\Delta^2$ into condensed and non-condensed
contributions
called $\Delta_{sc}^2$ and $\Delta_{pg}^2$ respectively. We will see
that the number of
non-condensed pairs associated with the pseudogap ($pg$) (represented by
$\Delta_{pg}^2$) can be determined once one knows their effective mass
$M^*$. And this, in turn, is determined by choosing a propagator for the
non-condensed pairs for which the BEC condition on the non-condensed
pair chemical potential, $\mu_{pair}=0$, is consistent with
Eq.~(\ref{eq:gap_equationmf}).

As expected, the BCS-Leggett approach, which 
is naturally associated with a T-matrix scheme, does not include
all the effects of Bogoliubov theory.
Within a T-matrix approach, one has a choice of factoring
$<c^\dagger c c^\dagger c>$ in one of two ways:
to yield either condensate terms
$\Delta_{sc}^2$ or pseudogap (pg) terms $\Delta_{pg}^2$.
At this level one
drops terms which couple the condensate
and pair excitations.
To mimic the effects of Bogoliubov-like theory, one would
need to introduce
cross terms of the form $\Delta_{sc}^2 \Delta_{pg}^2$ which
clearly involve higher order propagators and go beyond a
T-matrix approach.

By contrast the NSR-based approach uses Eq.~(\ref{eq:gap_equationmf})
with
\begin{equation}
\Delta_{mf} = \Delta_{sc} .
\end{equation}
That is, the ``gap'' parameter is replaced by the order parameter.
Equation~(\ref{eq:7a}) is not used. Rather one determines the fermionic
chemical potential $\mu^*$ by first establishing the bosonic
propagators. The latter are taken to be the collective mode propagators
for the $mf$ Hamiltonian but with the renormalized chemical potential
$\mu^*$.  The fermionic propagators, which also contribute to
determine $\mu^*$, are
derived via a T-matrix approach which couples the fermions and bosons.

From Eqs. (\ref{eq:new1}) and (\ref{eq:sum}) we see that in the BCS-Leggett based approach the
fermionic quasi-particle dispersion $\Ek$, which appears in the gap
equation, contains pseudogap effects. That is, the fermions which pair
are not the bare fermions. However the bosonic dispersion $\Omega_q^o$,
which also contributes to a separate (pseudo) gap equation, contains
interaction effects in a mean field sense only 
via a renormalized effective mass $M^*$. By contrast, the
NSR-based approach is based on a fermionic quasi-particle dispersion
$\Ek^0$ in which the fermions which pair are the bare fermions. However,
many body effects enter via a renormalized chemical potential $\mu^*$.
The interacting bosonic dispersion relation $\Omega_q = cq$ is derived.
Interestingly, the complexity of both approaches, at the level of 
numerical implementation, may lie in determining either renormalized
parameter $M^*$ or $\mu^*$, which, in a compact way,
reflects an approximate treatment of many body
effects in the respective theories.

\section{BCS-Leggett Approach at Finite $T \leq T_c$}
\label{sec:3}

\subsection{Theoretical Framework}
\label{sec:3f}

At issue then is the incorporation of bosonic degrees of freedom into
the gap and number equations.  The two different approaches build on the
fact that there are two different ways of arriving at soft bosonic modes
within a generalized BCS structure. These modes may come from the
collective phase mode of the \textit{order parameter} (Goldstone boson)
which is necessarily gapless in the superfluid phase. They may also
arise from the condition that the non-condensed pair excitation
spectrum is gapless.
Both of these are simultaneously satisfied in both
theory classes.

We turn first to the BCS-Leggett based theory, which provides a very
natural and straightforward extension of BCS theory. 
We note that strict BCS theory has two distinct conditions for 
soft modes of two particle propagators, 
one coming from the Goldstone boson and the other from
Eq.~(\ref{eq:5c}). This observation plays an important role in the
extension of BCS-Leggett theory to finite $T$. We begin by presenting
the central equations, rather than giving a
full derivation.
Two of these equations have already been written down
in Section \ref{sec:2c} for the superfluid regime. These are 
Eqs.~(\ref{eq:gap_equationmf})
and
Eq.~(\ref{eq:7a}),
importantly, with
the substitution 
$\Delta_{mf}(T) = \Delta(T)$, as in
Eq.~(\ref{eq:new1}). 

In order to quantify the pair
fluctuations, our task is to decompose $\Delta^2(T)$ into $\Delta_{sc}^2(T)$
and $\Delta_{pg}^2(T)$. 
The difference between the gap $\Delta$ and the order parameter $\Delta_{sc}$
is to be
associated with pair fluctuations (involving $\Delta_{pg}$), as should
be implicitly evident
in Eq.~(\ref{eq:sum}).
The physical arguments which we apply next are rather analogous to Bose
Einstein condensation: once we know the propagator for the non-condensed
pairs we determine the number of such pairs and in this way
we
determine $\Delta_{pg}^2$.
We can essentially anticipate the answer
simply by counting all non-condensed pairs as 
\begin{equation}
\Delta_{pg}^2 (T) = Z^{-1} \sum b(\Omega_q^o, T) ,
\label{eq:14}
\end{equation}
where $Z$ is an overall coefficient of proportionality, to be determined
below and $b(\omega,T)=1/[\exp(\omega/T)-1]$ is the Bose function. Here $\Omega_q^o$ is
the non-condensed pair dispersion.
Then just as in BEC theory, knowing the non-condensed pair
contribution ($\Delta_{pg}^2$) and the total ($\Delta^2$) one
can find the condensate term $\Delta_{sc}^2$.

To make progress we need to evaluate $\Omega_q^o$ (and $Z$). We equate
the
condition that the
propagator for non-condensed pairs has zero chemical potential
\begin{equation}
\mu_{pair} = 0
\label{eq:9}
\end{equation}
at and below $T_c$, with the gap equation
Eq.~(\ref{eq:gap_equationmf}), where
Eq.~(\ref{eq:new1})
must be imposed, so that we have 
\begin{equation}
1  + U  \mathop{\sum_{\bf k}}  \frac{1 - 2 f(E_{\bf k})}{2
E_{\bf k}}  = 0,
\qquad  T \le T_c
\label{eq:gap_equation}
\end{equation}
with 
\begin{equation}
\Ek \equiv \sqrt{ (\ek -\mu)^2 + \Delta^2(T)  } .
\label{eq:13}
\end{equation}
Note that it is the excitation gap and not the order parameter which
appears here.  That the BCS form for the gap equation is equivalent to
the gapless condition on non-condensed pairs imposes a constraint on
the non condensed pair propagator which must be of the form
\begin{equation}
t_{pg}(Q)= U/ [1+U \chi(Q)], 
\label{eq:14a}
\end{equation}
where, importantly, one must take the pair susceptibility
\begin{equation}
\chi(Q)=\sum_{K}G_{0}(Q-K)G(K) .
\label{eq:15}
\end{equation}
Here $G$ and $G_0$ are the full and bare Green's functions respectively.
We have met the combination $GG_0$ in the context of our
review of conventional BCS theory.  To expand on this point, note that
the full Green's function
is determined in terms of the usual BCS-like form for the self energy
\begin{equation}
\Sigma({\bf k}, \omega ) = \Delta^2 / [\omega
+\epsilon_{\bf k}-\mu  ]
\qquad  T \le T_c .
\label{eq:12}
\end{equation}
Using this self energy, one determines $G$ and thereby can evaluate
$t_{pg}$. The gap equation
in
Eq.~(\ref{eq:gap_equation})
thus requires that $t_{pg}(0) = \infty$.
Similarly, using
\begin{equation}
n = 2 \sum_K G(K)
\label{eq:22}
\end{equation}
one derives
\begin{equation}
n  = \sum _{\bf k} \left[ 1 -\frac{\ek - \mu}{\Ek}
+2\frac{\ek - \mu}{\Ek}f(\Ek)  \right]
\label{eq:23}
\end{equation}
which is the natural generalization of
Eq.~(\ref{eq:7a}).

The final set of equations which must be solved is rather simple and given by 
Eq.~(\ref{eq:23}), 
Eq.~(\ref{eq:14}), and 
Eq.~(\ref{eq:gap_equation}). This set 
has a more detailed derivation, and we summarize it
by noting that there there are two contributions to the full $T$-matrix
$t = t_{pg} + t_{sc}$ where $t_{sc}(Q)= -\frac{\Delta_{sc}^2}{T}
\delta(Q)$. 
Similarly, we have for the fermion self energy $\Sigma(K) =
\Sigma_{sc}(K) + \Sigma_{pg} (K) = \sum_Q t(Q)G_{0} (Q-K).$
It follows then that
\begin{equation}
  \Sigma_{sc}(\mb{k},\omega)=
  \frac{\Delta_{\mb{k},sc}^2}{\omega +\ek-\mu} .
\label{SigmaSC}
\end{equation}

A vanishing chemical potential means that $t_{pg}(Q)$ diverges at $Q=0$
when $T\le T_c$. Thus, we approximate \cite{Maly1,Kosztin1} $\Sigma(K)$
to yield
\begin{equation}
\Sigma_{pg} (K)\approx -G_{0} (-K) \Delta_{pg}^2 ~~~T \leq T_c\,,
\label{eq:sigma3}
\end{equation}
with
\begin{equation}
\Delta_{pg}^2 \equiv -\sum_{Q\neq 0} t_{pg}(Q). 
\label{eq:18}
\end{equation}
This equation will be shown below to be equivalent to Eq.~(\ref{eq:14}).
We write
\begin{equation}
\Sigma_{pg}(\mb{k},\omega) \approx 
\frac{\Delta_{\mb{k},pg}^2}{\omega+\ek-\mu} 
\label{SigmaPG_Model}
\end{equation}
from which 
one finds
$\Sigma({\bf k}, \omega ) \approx \Delta^2 / [\omega
+\epsilon_{\bf k}-\mu  ]$ where
we have used Eq.~(\ref{eq:sum}). In this way one derives
Eq.~(\ref{eq:gap_equation}).

Note that in the normal state (where $\mu_{pair}$ is nonzero),
Eq.~(\ref{eq:sigma3}) is no longer a good approximation, although
a natural extension can be readily written down \cite{heyan2}.

At small four-vector $Q$, we may expand the inverse of $t_{pg}$ after
analytical continuation. Because we are interested in the
moderate and strong coupling cases, where the contribution of the quadratic term in $\Omega$ term is small, we drop this term and thus find the following expression, which 
yields 
$\Omega^{o}_q = q^2/(2M^*)$ via the expansion 
\begin{equation}
t_{pg}(Q) = \frac { Z^{-1}}{\Omega - \Omega^{o}_q +\mu_{pair} + i \Gamma^{}_Q},
\label{eq:24}
\end{equation}
where $Z$ is a residue given by 
\begin{eqnarray}
  Z&=&\frac{\partial t_{pg}^{-1}}{\partial\Omega}\Big|_{\Omega=0,q=0} \nonumber \\
  &=&\frac{1}{2\Delta^2}\left[n-2\sum_{\mathbf{k}}f(\ek-\mu)\right].
\label{eq:25}
\end{eqnarray}
Further details are presented in 
Appendix \ref{App:A}.

Below $T_c$ the imaginary contribution in Eq. (\ref{eq:24})
$\Gamma^{}_Q \rightarrow 0$
faster than $q^2$ as $q\rightarrow 0$.
It should be stressed that this approach yields the ground state
equations and that it represents a physically meaningful extension of
this ground state to finite $T$.

We note that the approximation in
Eq.~(\ref{eq:sigma3})
is not central to the physics, but it does greatly simplify the
numerical analysis. One can see that correlations which do not
involve pairing, such as Hartree terms are not included here.
This is what is required to arrive at the BCS-Leggett ground state.
It should be clear that, in principle, the T-matrix approach
discussed here is more general and that 
in order to address experiments at a more quantitative level
it will be necessary to go beyond 
Eq.~(\ref{eq:sigma3}). Indeed, the simplest phenomenological
correction is to write
\begin{equation}
\Sigma_{pg}(\mb{k},\omega) \approx 
\frac{\Delta_{\mb{k},pg}^2}{\omega+\ek-\mu+i\gamma} +\Sigma_0 (\mb{k},\omega) .
\label{SigmaPG_Model_Eq}
\end{equation}
Here the broadening $\gamma \ne 0$ and ``incoherent'' background
contribution $\Sigma_0$ reflect the fact that noncondensed pairs do not lead to
\textit{true} off-diagonal long-range order.
By contrast $\Sigma_{sc}$ is associated with long-lived
condensed Cooper pairs, and as shown in 
Eq.~(\ref{SigmaSC}), it is 
similar to $\Sigma_{pg}$ but without the broadening.
It is important to note that this same analysis has been applied to
describing the spectral function
in the pseudogap \cite{Normanarcs,Chubukov2} and the superfluid
phases \cite{FermiArcs} of the high temperature superconductors, where
here $\Sigma_0(\mb{k},\omega)$ is taken to be an imaginary constant.

In summary, the simplifying approximation in Eq. (\ref{eq:sigma3}) is most problematic
when the pairing gap is small so that other correlations and contributions
(which are otherwise in the ``background") become important. Perhaps the
most nobable example of when this simplification affects the
qualitative physics is in the population imbalanced gases. At a quantitative
level, a clear shortcoming comes from the neglect of Hartree interaction
effects. These issues are discussed in Section \ref{sec:5a}.

Finally, we present results for the thermodynamical potential , which is
given by
\begin{eqnarray}
\Omega &=& \Omega_f + \Omega_b, \nonumber\\
\Omega_f&=&\Delta^2\chi(0)+\sumk[(\ek-\mu-\Ek)-2T\ln(1+e^{-\Ek/T})],\nonumber\\
\Omega_b&=&\sumq T\ln(1-e^{-\Omega^{o}_q/T}). 
\label{eq:28}
\end{eqnarray}
This thermodynamical potential can be used to generate the self
consistent equations presented above
\begin{equation}
\frac{\p\Omega}{\p\Delta}=0
\label{eq:27}
\end{equation}
which is equivalent to the gap equation of Eq.~(\ref{eq:gap_equation})
Similarly, we have
\begin{equation}
\frac{\p\Omega}{\p\mu_{pair}}=0
\label{eq:28a}
\end{equation}
which leads to the equation for the pseudogap given by
Eq.~(\ref{eq:18}).  Finally, the number equation
\begin{equation}
n=-\frac{\p\Omega}{\p\mu}
\label{eq:29}
\end{equation}
which yields Eq.~(\ref{eq:23}).

We recapitulate by rewriting the central gapless condition for the
non-condensed pairs as
\begin{equation}
t_{pg}(0) = \frac {U} {1 + U \sum_{\bf k} G_0 (-K) G(K) } = \infty .
\label{eq:30}
\end{equation}
This equation is equivalent to Eq.~(\ref{eq:gap_equation}) or
Eq.~(\ref{eq:9}).  Expanding $t_{pg}(Q)$ determines the excited
pair dispersion
\begin{equation}
\Omega_q^o = q^2/2M^*. 
\label{eq:31}
\end{equation}

\subsection{BCS-Leggett Approach to BEC}
\label{sec:3a}

There has been some confusion voiced about whether the BCS-Leggett
ground state requires that one ignore 
bosonic degrees of freedom. To respond (in the negative)
to this concern it is useful to address the extreme BEC limit.
We begin by making the important observation \cite{JS3}
that for $ T \leq T_c$, the
fermionic parameters associated with the wavefunction of
Eq.~(\ref{eq:1a}), \textit{$\Delta(T)$ and $\mu(T)$ are temperature
  independent in the BEC, for all $ T \leq T_c$}.  
This is consistent with the physical picture of well established,
pre-formed pairs in the BEC limit, so that the fermionic energy scales
are unaffected by $T$ below $T_c$.

We now
extend these qualitative observations to a more quantitative level.  The
self consistent equations in the BEC limit for general temperature $T$
can then be written
as
\begin{eqnarray}
\frac{m}{4 \pi \hbar^2 a } & = & \mathop{\sum_{\bf k}}\left[ \frac{1}{2
 \ek } - \frac{1}{2E_{\bf k}} \right]  \,, \label{eq:3a}\\
 n & =& \sum _{\bf k} \left[ 1 -\frac{\ek - \mu}{\Ek}
  \right] \,, ~~~~T \leq T_c ,
 \label{eq:4a}
 \end{eqnarray}
 where we have now introduced the usual s-wave scattering length, $a$,
 which is needed 
 to regularize the gap equation for a contact
 potential.
 Note that we have used the $T=0$ conditions \cite{Leggett} in
 Eqs.~(\ref{eq:3a}) and (\ref{eq:4a}), since the Fermi function $f(\Ek)$
 is essentially zero in the BEC limit, where $\Ek/T \gg 1$.
%
 Equations ~(\ref{eq:3a}) and (\ref{eq:4a}) are central to the
 BEC-theory.  They show that even in the strong attraction limit, where
 the system can be viewed as consisting of ``bosons'', \textit{the
   underlying fermionic constraints on $\Delta$ and $\mu$ must be
   respected}.

 It follows from the above equations that
for general $ T \le T_c$,
 \begin{equation}
 n_{pairs} \equiv \frac{n}{2}= Z \Delta^2 ,
 \label{eq:11}
 \end{equation}
where the coefficient of proportionality  
\begin{equation}
Z \approx \frac {m^2 a } { 8 \pi \hbar^4} .
\label{eq:10}
\end{equation}
This coefficient $Z$ was obtained directly
from the ground state equations
\cite{Strinati,Stringari}. However, it can also be readily
derived at non-zero $T$
using the propagator for non-condensed pairs following
Eq.~(\ref{eq:25}). Here one drops the last
term involving the summation over free fermion states, which are clearly
negligible in the BEC.
That the same answer is obtained from the ground state and from
$t_{pg}(Q) $ demonstrates an internal consistency of the calculations.

We arrive at an important physical interpretation of the BEC limit.
Even though $\Delta$ or $n_{pairs}$ is a constant in $T$, this constant
must be the sum of two temperature dependent terms.  Indeed it follows
from Eq.~(\ref{eq:sum}) that, just as in the usual theory of BEC these
two contributions correspond to condensed and non-condensed components
\begin{equation}
\frac{n}{2}=  n_{pairs}^{condensed}(T) + n_{pairs} ^ {noncondensed}(T) .
 \label{eq:12a}
 \end{equation}
Note also that at $T_c$
 \begin{equation}
 n_{pairs}^{noncondensed}(T_c) = \frac{n}{2}= \sum_{\bf q} b
 (\Omega_q^o,T_c) .
 \label{eq:52}
 \end{equation}

 We now rewrite the central equations (\ref{eq:3a}), (\ref{eq:4a})
 in the BEC limit to compare more directly with the case of a
 weakly interacting Bose gas.
\begin{equation}
n =  \Delta^2 \frac{  m^2} {4 \pi \sqrt{2m |\mu|} \hbar^3},
\label{eq:7}
\end{equation}
which, in conjunction with the expansion of Eq.~(\ref{eq:3a}),
  
\begin{equation}
\frac{m}{4 \pi \hbar^2 a}=\left(\frac{2 m}{\hbar^2}\right)^{3/2}
\frac{\sqrt{|\mu|}}{8 \pi}
\left[1+\frac{1}{16} \frac{\Delta^2}{\mu^2}\right],
\label{eq:gap_eq_exp}
\end{equation}
yields
%
\begin{equation}
\mu = - \frac{\hbar^2}{2 m a^2} + \frac{a \pi n \hbar^2} {m}.
\label{eq:mu_Uonly}
\end{equation}

These expressions are used to eliminate the fermionic parameters
altogether and arrive at an expression which, at $T=0$ some have
interpreted \cite{Stringari,Strinati} to be equivalent to the results of
Gross Pitaevski (GP) theory. Here one identifies an effective inter-pair
scattering length $a_B \equiv 2a$
with $n_B \equiv n/2$ which represents the number density of pairs, and
finally $\mu_B \equiv 2\mu + \hbar^2/ma^2$ is the ``bare'' chemical
potential of the pairs, with $M_B\equiv 2m$ the pair mass.  We emphasize
that the value of 2 for the scattering length ratio is entirely dictated
by the assumed form for the ground state, Eq.~(\ref{eq:1a}).

With these definitions,

\begin{equation}
\mu_B  = \frac {4 \pi a_B \hbar^2} {M_B} 
(n) .
\label{eq:43}
\end{equation}
For true bosonic
systems, this GP equation is usually considered only at $T=0$, where all
the pairs are condensed. In this regard we should interpret $\mu_B$
as a ``bare" chemical potential
which includes only a mean field Hartree shift. This is to be
contrasted with
$\mu_{pair}$ which is the
chemical potential of the non-condensed pairs and reflects
many body physics beyond Hartree terms.
Similarly $M^*$ is the effective mass of the non-condensed pairs which
is generally distinct from $M_B$.

Note, however, that our derivation of Eq.~(\ref{eq:43}) should, in
principle, apply to all $ T \leq T_c$, and, thus, the physics is very
different from that of GP theory.  Clearly $\mu_B $ as defined above is
a constant in temperature. The quantity $n$ appearing in
Eq.~(\ref{eq:43}) is, of course, temperature independent, but we note
here that via Eq.~(\ref{eq:12a}) it contains both condensed and
non-condensed pairs.  Their relative contribution can be determined via
an ideal gas dispersion relation with renormalized effective mass. This
$\Omega_q ^o \propto q^2$ dispersion is, in turn, a consequence of the
underlying gap equation Eq.~(\ref{eq:3a}).  We stress that this gap
equation has no counterpart in the GP theory for true bosons, although
it can be interpreted in the fermionic context as reflecting the
condition that $\mu_{pair} =0$.

Another essential distinction between the fermionic BEC and that of true
bosons is that the effective mass contains interaction effects due to
compositeness.  The general expression for the (non-condensed) pair mass
$1/M^*$ in the near BEC limit is given by
\begin{equation}
\frac{1}{M^*}=\frac{1}{Z \Delta^2} \sum_{\bf k}
\left[ \frac{1}{ m} \vk^2
- \frac{4 \Ek \hbar^2 k^2}{3m^2 \Delta^2} \vk^4\right] ,
\label{eq:B}
\end{equation}
where we have used 
Eq.~(\ref{eq:24}), as well as Eq.~(\ref{eq:3a}) and (\ref{eq:4a}).
After expanding to lowest order in $na^3$,
\begin{equation}
M^* \approx 2m \left( 1 + \frac{\pi a^3 n}{2}\right) .
\label{eq:mass_change}
\end{equation}
Physically this increase in effective mass away from the ideal gas
asymptote reflects the fact that pairs are less mobile, as a consequence
of the inter-pair repulsion. This means
that the asymptotic limit of $T_c$ is approached \textit{from below},
which is different from the behavior found in the NSR approach
\cite{NSR}.
The issue of whether the asymptotic limit for $T_c$ in 
a mean field composite BEC should
be approached from above or below has been addressed \cite{Haussmann}
in the literature, where it was argued in favor of the latter alternative.

We turn now to a quantitative calculation of $T_c$, based on
$\Omega_q^o$ [via Eq.~(\ref{eq:52})].  Equation~(\ref{eq:52}) reflects
the fact that, in the near-BEC limit, and at $T_c$, all fermions are
constituents of uncondensed pairs.  It, then, follows that $(M^*
T_c)^{3/2} \propto n = const.$ which, in conjunction with
Eq.~(\ref{eq:mass_change}) implies
\begin{equation}
\frac{ T_c - T_c^0}{T_c^0} = - \frac{\pi a^3 n}{2}.
\label{eq:46}
\end{equation}
Here $T_c^0$ is the transition temperature of the ideal Bose gas with
$M_B=2m$. This downward shift of $T_c$ follows the effective mass
renormalization, much as expected in a Hartree treatment of GP theory at
$T_c$. Here, however, in contrast to GP theory for a homogeneous system
with a contact potential \cite{RMP}, there is a non-vanishing
renormalization of the effective mass.

\subsection{Bogoliubov de Gennes Theory and Critical Velocity Calculations}
\label{sec:3b}

The most widely used theoretical formalism for the trapped Bose gases is
probably Gross Pitaevski theory \cite{RMP}.  This is because it has the
flexibility to address inhomogeneous systems and general perturbations.
For the trapped Fermi gases, the emerging counterpart formalism appears to be
Bogoliubov de Gennes (BdG) theory.  Both BdG and GP theory are presumed
to be appropriate to the ground state.  Moreover the ground state in
question for the Fermi gases is associated with the BCS-Leggett
wavefunction.

The BdG equation is
\begin{equation}
\left(\begin{array}{cc}
H(\mathbf{r}) & \Delta(\mathbf{r}) \\
\Delta(\mathbf{r}) & -H(\mathbf{r})
\end{array}\right)\left(\begin{array}{c}
u_n(\mathbf{r}) \\
v_n(\mathbf{r})
\end{array}\right)=E_n\left(\begin{array}{c}
u_n(\mathbf{r}) \\
v_n(\mathbf{r})
\end{array}\right).
\end{equation}
Here $H(\mathbf{r})=\frac{\hbar^{2}}{2m}\nabla^2-\mu$.  
The solution of these equations is subject to the self consistent
gap and number equations
\begin{equation}\label{eq:Delta}
\Delta(\mathbf{r})=-U\sum_n u_n(\mathbf{r})v_n^*(\mathbf{r})[1-2f(E_n)] 
\end{equation}
and
\begin{equation}\label{eq:n}
  n(\mathbf{r})=\sum_{\sigma,n}\left\{|u_n(\mathbf{r})|^2f(E_n)+|v_n(\mathbf{r})|^2[1-2f(E_n)]\right\}.
\end{equation}
Finally,
the mass current is
\begin{equation}\label{eq:J}
\mathbf{J}(\mathbf{r})=2\{\mathbf{J}_{u_n}f(E_n)-\mathbf{J}_{v_n}[1-f(E_n)]\} ,
\end{equation}
%
where $\mathbf{J}_{u_n}=\mbox{Im}(u_n^*\nabla u_n)$ and
$\mathbf{J}_{v_n}=\mbox{Im}(v_n^*\nabla v_n)$.  The general solution to
the BdG equation depends on the geometries and coupling constant and
therefore usually requires full numerical calculation.

This system of equations has been applied to the problem of BCS-BEC
crossover in a number of important ways at $T=0$. It was shown in
Reference \cite{PSP03} that in the deep BEC, this scheme becomes
equivalent to Gross Pitaevski theory.  This observation may not be, in
some sense, entirely surprising based on the arguments we have just
presented in 
Section \ref{sec:3a}.
Moreover, one can see that there is a close
analogy between the wavefunction of Eq.~(\ref{eq:1a}) and that of Gross
Pitaevski theory for point bosons.  BdG theory has been used to address
the behavior of a single vortex \cite{Machida,Chienvortex,Hovortex} as
the system evolves from BCS to BEC. One of the key observations here is
that the core size (related to the coherence length $\xi$) is
non-monotonic with scattering length, exhibiting a minimum near
unitarity.  Moreover, there have been systematic studies based on BdG
theory in the presence of \cite{BdG_Imbalance} population imbalance.
Finally, we want to call attention to work which addresses the
behavior of the critical current as extracted from both vortex
calculations \cite{Hovortex} and from Josephson junction studies
\cite{StrinatiJosephson}. Direct calculations using Eq.~(\ref{eq:J})
show a maximum in this current as a function of distance from the vortex
core center and this maximum can be loosely associated with the critical
current, $I_c$. Because $I_c$ scales inversely with $\xi$, one can infer
from BdG calculations of $\xi$ \cite{Chienvortex} 
that the critical current is largest close to unitarity, as observed
experimentally \cite{Ketterle_Criticalcurrent}.

Physically, this maximum in $I_c$ has been interpreted
\cite{Stringaricv,Hovortex} as suggesting that on the BCS side of
resonance $I_c$ is determined by the breaking of \textit{condensate}
pairs, while on the BEC side of resonance, $I_c$ reflects the collective
modes of the condensate.  These two different mechanisms have different
dependences on the fermionic scattering length, leading to a maximum
which one might argue is close to unitarity.  We emphasize here that
$I_c$ is a property of the condensed pairs within BCS-Leggett theory. As
noted earlier, one has to exercise caution in applying the so-called
Landau criterion in calculating $I_c$. Only those excitations 
which couple to the
condensate (that is, to the density) are to be included in establishing
the stability of the superfluid.

\subsection{Superfluid Density and Collective Mode Calculations}
\label{sec:3c}

We noted in 
Section \ref{sec:2b},
that the superfluid density $n_s$ is highly
constrained by a Ward identity once the self energy is chosen.  These
considerations have been applied \cite{Kosztin1,Chen2} to the
BCS-Leggett-based formalism where it has been shown that the
contribution of non-condensed pairs does not directly contribute to a
Meissner effect, as expected. The Aslamazov-Larkin and Maki Thompson
diagrams associated with these finite momentum pairs cancel out and one
is left with only a condensate contribution of the form

\begin{equation}
\left( \frac{n_s}{m} \right)  = \frac{\Delta_{sc}^2}{\Delta^2}
\left ( \frac{n_s}{m} \right)^{BCS}  \:,
\label{Lambda_BCS_Eq}
\end{equation}
where $(n_s/m)^{BCS}$ is defined in
Eq.~(\ref{eq:nsBCS}), but with $\Delta_{sc}$ now replaced
by $\Delta$. Obviously, $(n_s/m) ^{BCS}$ does not vanish at
$T_c$, but because of the prefactor, the superfluid density reflects the
order parameter and will be zero in the normal state.  One can interpret
this expression 
using
$\Delta_{sc}^2
(T) = \Delta^2 (T) - \Delta_{pg}^2(T)$.
and
noting that there are two forms of condensate
excitation which lead to a decrease in superfluid density with
increasing $T$; the fermionic excitations, which are important to the
extent that $\Delta(T)$ contains an appreciable temperature dependence
below $T_c$, and the non-condensed pairs which enter via $\Delta_{pg}^2
(T)$.

In a related fashion, there is an extensive literature
\cite{Cote,Randeria2,Kosztin2} which has addressed the $T=0$ collective
modes of the BCS-Leggett state.  In the BCS limit the sound mode
velocity is $c(T=0) = v_F / \sqrt{3}$, while in the BEC limit $ c(T=0)
\approx \sqrt {(4 \pi n a_B / M_B^2)}$,  with the inter-boson scattering
length $a_B = 2 a$, as derived in 
Section \ref{sec:3b}. 
As noted in 
Section \ref{sec:2b}, 
the
inter-boson interactions arise in the condensate dynamics just as in
Eq.~(\ref{eq:response}) through the presence of four Green's functions
in the second term in this expression.

With the introduction of non-zero temperature, the collective mode
spectrum must be deduced on the basis of a gauge invariant formulation
of the response of the system to a fictitious vector potential,
which enforces the Ward
identity constraints deriving from the self energy.
Because $\Delta(T) \neq \Delta_{sc}(T)$, this calculation is much more
difficult to implement. A lowest order approximation was discussed in
Ref. \cite{Kosztin2}.  In this case $c(T)$ becomes complex, but
both real and imaginary contributions are seen to vanish at $T_c$.  If
there is to be an eventual reconciliation between the two approaches to
BCS-BEC crossover it will be necessary, at the least to find a full
solution to this problem.

\section{Nozieres Schmitt-Rink theory:
Bogoliubov-based Approach to finite $T \leq T_c$}
\label{sec:4}

Although the normal state is similar to that originally proposed by
Nozieres and Schmitt-Rink, the
philosophy underlying this theoretical scheme for describing
BCS-BEC crossover \cite{StrinatiPRB70} 
begins with Galitskii's approach \cite{Fetter} to the dilute Fermi
gas with repulsive interactions. Here a self-energy based on a particle-particle
ladder is introduced. Moreover, it is clear that this scheme
can be readily extended to
the case of a weak attractive interaction in the normal phase, and then
further extrapolated to
the BEC limit (still
remaining in the normal phase), where the particle-particle ladder acquires
the form of the propagator for non-interacting bosons.
It then becomes natural to extend this scheme
to the superfluid
phase, for which the particle-particle ladder acquires a matrix structure
that maps onto the bosonic normal and anomalous propagators within 
Bogoliubov theory. For these reasons the main physical emphasis was on the
self-energy itself, and as a consequence on the related dynamical quantities.

One of the virtues of this type of diagrammatic
approach is that it is "modular" in nature, in the sense that it can be
progressively improved by including additional self-energy corrections which
are thought to be important, particularly at the BCS and BEC endpoints.
In this way, upon successive improvements  one can address
the Popov theory for composite bosons, 
the Gorkov and Melik-Melik-Barkudarov corrections \cite{StrinatiPRB70}, etc. Of
course, practical implementation of these theoretical improvements suffers by
the increased numerical complexity.

For want of a better short name, we will refer to this as the ``NSR-based
approach". One can also think of it as a diagrammatic T-matrix scheme which involves a 
\textit{matrix} form of the T-matrix. By contrast the BCS-Leggett approach
is a diagrammatic T-matrix scheme which involves a \textit{scalar} form
for the T-matrix.

In this alternative approach,
Eq.~(\ref{eq:gap_equationmf}) is used \cite{PS05,Griffin2} to yield
\begin{equation}
1  + U  \mathop{\sum_{\bf k}}  \frac{1 - 2 f(E_{\bf k}^o)}{2
E_{\bf k}^o}  = 0,
\qquad  T \le T_c
\label{eq:gap_equationNSR}
\end{equation}
with
\begin{equation}
\Ek^o \equiv \sqrt{ (\ek -\mu^*)^2 + \Delta_{sc}^2 (T) } .
\label{eq:dispersionNSR}
\end{equation}

We can rewrite this gap equation, along with the number equation as
\begin{eqnarray} 
\Delta_{sc}&=&-U\sum_K \hat{G}_{12}^{o}(K)\label{eq:NSR49} \\
n&=&2\sum_K \hat{G}_{11}(K).
\label{eq:NSR50}
\end{eqnarray} 
Here $\hat{G}^{o}$ is the bare \textit{matrix} Green function with
components given by
$\hat{G}_{11}^{o}=-(\xik+i\omega_n)/[(\Ekk)^2+\omega_n^2]$ and
$\hat{G}_{12}^{o}=\Delta_{sc}/[(\Ekk)^2+\omega_n^2]$ with
$\xik=k^2/2m-\mu^*$.  Note that there are two different levels of
Green's functions which appear in these equations.  In effect,
fluctuations associated with the collective modes will appear in the
number equation, but not the gap equation.

The fully dressed Green's functions which include collective mode
effects are determined in terms of the matrix self energies 
\ba
& &\Sigma_{11}(\vvk,\omega_n)=-\Sigma_{22}(-\vvk,-\omega_n)\nonumber\\
&=&-\sum_{Q}\Gamma_{11}(Q) \hat{G}_{11}^{o}(Q -K), \\
& &\Sigma_{12}(\vvk,\omega_n)=\Sigma_{21}(\vvk,\omega_n)=-\Delta_{sc} ,
\ea 
from which the important dressed Green's function
(which reflects) the pair fluctuations and which is used in the number
equation Eq.~(\ref{eq:NSR50}) can be derived  \cite{StrinatiPRB70}:
\begin{eqnarray}
\hat{G}_{11}(K) &=& \frac {1}{G_{0}^{-1}(K) - \sigma_{11}(K)}, \\
\sigma_{11}(K) &=& \Sigma_{11}(K)+\frac{\Sigma_{12}(K)\Sigma_{21}(K)}{G_{0}^{-1}(K)-\Sigma_{22}(K)}.\nonumber
\label{eq:53}
\end{eqnarray}
Here $G_{0}^{-1}(K)=(i\omega_n-\xi_{\bf k})$.

The pair propagator (which is the analogue of $t_{pg}$ in the
BCS-Leggett theory) is related to the ``bare collective modes''.
In particular,
$$\Gamma_{11}(Q)=\frac{\chi_{11}^0(-Q)}{\chi_{11}^0(Q) \chi_{11}^0(-Q)-[\chi_{12}^0(Q)]^2}$$
\begin{eqnarray}
  -\chi_{11}^0(Q)&=&\sum_{K}\hat{G}_{11}^{o}(K+Q)\hat{G}_{11}^{o}(-K)-\frac{1}{U}\\
  \chi_{12}^0(Q)&=&\sum_{K}\hat{G}_{12}^{o}(K+Q)\hat{G}_{21}^{o}(-K),
\label{eq:Collective}
\end{eqnarray} 
where throughout we use the four vector notation $K=(\vvk,\omega_n)$ and
$Q=(\vq,\Omega_{\nu})$.

We end by recapitulating the central gapless condition of this class of
theories:
\begin{equation}
\Gamma_{11}(0)=\frac{\chi_{11}^0(0)}{\chi_{11}^0(0)\chi_{11}^0(-0)-[\chi_{12}^0(0)]^2}
= \infty .
\label{eq:NSRgapless}
\end{equation}

The speed of sound is obtained from
the
finite $Q$ generalization of the denominator in
Eq.~(\ref{eq:NSRgapless}): 
$ \chi_{11}^0(Q) \chi_{11}^0(-Q)-[\chi_{12
}^0(Q)]^2 =0$
which can be seen \cite{Stringaricv}
to yield an answer equivalent to that obtained from
Eq.~(\ref{eq:coll}).
Quite generally at small
wave-vector this bosonic dispersion is given by
\begin{equation}
\Omega_q = c(T)  q.
\label{eq:57}
\end{equation}

\subsection{Nozieres Schmitt-Rink-Based Theory in the BEC limit:
$ T \leq T_c$}
\label{sec:4a}

It has been shown that \cite{PS05} in the BEC limit, the equations for
the collective mode propagators $\Gamma_{11}$ and $\Gamma_{12}$ are very
similar to the diagonal and off-diagonal bosonic Green's functions at
the level of Bogoliubov theory \cite{Griffin4}.  In the deep BEC these
bosonic Green's functions have a pole at
\begin{equation}
\Omega_{\bf q } = \sqrt {  \left( \frac {{\bf q}^2} {2 M_B} + \mu_B \right) ^2
- \mu_B ^2 }
\label{eq:58}
\end{equation}
which represents the characteristic dispersion relation for bosons in a
weakly interacting Bose gas.  Here $\mu_B $
is defined in Eq (\ref{eq:43}) and $M_B = 2m $ is
the boson mass.

The associated fermionic Green's functions are in some sense the more
important, since these are fundamentally fermionic gases.  In the BEC
limit the equation
\begin{equation}
\Sigma_{11}(\vvk,\omega_n)
=-\sum_{Q}\Gamma_{11}(Q) \hat{G}_{11}^{o}(Q -K) 
\label{eq:59}
\end{equation}
can be approximated by ignoring terms which involve $\Delta_{sc}$
compared to $|\mu^*|$. The fermion Green's functions in the BEC limit are approximated as the following expressions which are derived in Ref.~\cite{PS05} and we summarize the derivation in Appendix \ref{App:B}.
\begin{equation}
\hat{G}_{11}=-(\xik+i\omega_n)/[\omega_n^2 + \xik^2 + \bar{\Delta}_{pg}^2 + \Delta_{sc}^2] .
\label{eq:60}
\end{equation}
Here the approximation
\begin{equation}
  \bar{\Delta}_{pg}^2  \approx    -\sum_{Q}\Gamma_{11}(Q)
\label{eq:61}
\end{equation}
has been used. This approximation is similar in spirit
to that shown in Eq.~(\ref{eq:sigma3}).
It is also demonstrated \cite{PS05} that in the deep BEC regime, the fermion Green's function leads to
\begin{equation}
\sum_{K}\hat{G}_{11}(K)=\frac{n}{2}\approx n_0+n^{\prime}.
\end{equation}
Here $n_0$ and $n^{\prime}$ denote densities of condensed and
noncondensed pairs, respectively.  Similarly, it follows that
\begin{equation}
\hat{G}_{12}= \Delta_{sc} /[\omega_n^2 + \xik^2 + 2\bar{\Delta}_{pg}^2 + \Delta_{sc}^2] .
\label{eq:62}
\end{equation}
The modified gap equation,
\begin{equation}\label{eq:Strinatigap}
\Delta_{sc}=U\sum_{K}\hat{G}_{12}(K),
\end{equation}
gives \cite{PS05}, in the deep BEC limit, 
\begin{equation}
\mu_{B}\approx \left(\frac{4\pi a_{B}}{M_{B}}\right)(n_0+2n^{\prime}).
\end{equation}
Under these approximations, pairs in the deep BEC
limit behave like bosons in the Popov approximation.  Although there is
an asymmetry between the denominators of these two component Green's
functions, one can see that the diagonal term has a strong similarity to
the previous approach of 
Section \ref{sec:3f}. 
The effective excitation gap is given
by the contribution from condensed and excited pairs.

To go beyond this scheme, it is necessary to incorporate corrections to
the gap equation 
Eq.~(\ref{eq:gap_equationNSR})
with concomitantly
those to the collective mode spectrum,
so that collective mode effects have to
be treated at a level beyond the bare modes of BCS
theory. Some progress has been made \cite{PS05} in implementing this
scheme in the BEC limit.

\subsection{The Controversy Surrounding the Number Equation in The NSR Approach}
\label{sec:4b}

In the original NSR approach the number equation was determined from a
thermodynamical potential, Here, above and below \cite{Griffin2} $T_c$,
one approximates the thermodynamical potential
\begin{eqnarray}
\Omega_{NSR} &=&\Omega_{mf}+\Omega_{pf}^0, \label{eq:72NSR} \\
\Omega_{mf}&=&\sumk(\xik-\Ekk+\frac{\Delta_{sc}^2}{2k^2})\nonumber\\
&-&2T\sumk\ln(1+e^{-\Ekk/T}),\\
\Omega_{pf}^0&=&\sum_{Q}\ln[\chi_{11}^0(Q)\chi_{11}^0(-Q)-[\chi_{12}^0(Q)]^2] .
\label{eq:74a}
\end{eqnarray}
Note that in this RPA-like scheme, only the bare pair susceptibilities
$\chi^0$ are included.

Quite generally, the number equation is
given by
\begin{equation}
n = \frac{\p\Omega}{\p\mu^*}
\label{eq:75a}
\end{equation}
which is necessarily equivalent to
Eq.~(\ref{eq:NSR50}), providing one has a complete
theory.
However,
because of the lack of full self consistency,
the original NSR approach was criticized by Serene \cite{Serene}.
By
approximating the pair fluctuation contributions, it corresponds to a
T-matrix theory in which one takes only the lowest order terms in a
Dyson expansion, rather than a full resummation, so that
\begin{equation}
G(K) \approx  G_0(K) + G_0(K) \Sigma_0 (K) G_0(K) .
\label{eq:12f}
\end{equation}
This criticism, not withstanding, it
has recently been argued \cite{Drummond2,Drummond3,Randeriaab} that,
below $T_c$ one should write the number equation as
\begin{equation}
n=-\frac{d\Omega_{NSR}}{d\mu^*}=-\left(\frac{\p\Omega_{NSR}}{\p\mu^*}+\frac{\p\Omega_{NSR}}{\p\Delta_{s
c
}}\frac{d\Delta_{sc}}{d\mu^*}\right)
\label{eq:75b}
\end{equation}
with $d\Delta_{sc}/d\mu^*$ determined from the BCS gap equation.

In earlier work \cite{Griffin2}, the second term on the right hand side
of Eq(\ref{eq:75b}) was dropped.  Indeed the fact that
\begin{equation}
\frac{\p\Omega_{NSR}}{\p\Delta_{sc
}} \neq 0
\end{equation}
in our view reflects a problem in the theory-- that the gap equation is
non-variational, or non-self consistent.  
This non-variational behavior implies that a Landau Ginsburg like
analysis, and even its generalization to first order phase transitions,
is not possible.
This anomalous term appears
discontinuously below $T_c$ and, it will enhance first order
discontinuities at $T_c$, which may already be present in Bogoliubov or
Popov level approaches.

Nevertheless, it has been argued that in the BEC this non-variational
term provides a quantitative improvement over previous work since it
evidently yields the nearly correct \cite{Drummond3} relationship between
the inter-boson ($a_B$) and inter-fermion ($a$) scattering
lengths. Exact few body calculations \cite{Petrov} show that this ratio
should be $0.6$. It appears difficult to understand physically how an
evidently non-self consistent gap equation can capture the same physics
as these precise four fermion calculations.  Indeed, this claim appears
to be at odds with detailed calculations presented elsewhere which show
that to arrive at this correct ratio, one must go beyond
\cite{Kagan,PS05} T-matrix based schemes.

For ease in identification of these two different versions of NSR
theory, we now refer to that based on
Eq.~(\ref{eq:NSR50}) as NSR-1 and that based on Eq.~(\ref{eq:75b}) as
NSR-2.



\subsection{Superfluid Density and Collective Mode Calculations}
\label{sec:4c}

The superfluid density $n_s$ as a function of temperature has been
calculated using both NSR-1 and NSR-2 like theories.  For the former, a
diagrammatic calculation of the current-current correlation function
based on Aslamazov-Larkin and Maki-Thompson contributions was adopted
\cite{Strinati6}, which is, in many ways, similar to that discussed in
Section \ref{sec:3c}
within the BCS-Leggett framework \cite{Kosztin1,Chen2}.  For
NSR-2 like theories a framework based on changes in the thermodynamic
potential associated with a ``phase twist'' was adopted
\cite{Griffingroup1,Griffingroup2}.  The results appear to be rather
similar, at a qualitative level.  For some parameter regimes, there are
either first order transitions at $T_c$ or multivalued results for $n_s$
which presumably reflect the analogous behavior found in Bogoliubov or
Popov level treatments of true Bose systems \cite{AndersenRMP,Griffin4}.
See Appendix \ref{App:C}.

An important check on these calculations is to verify that there is no
Meissner effect in the normal state.  We can follow the same analysis as
used in Eq.~(\ref{eq:kernel}). Quite generally, above $T_c$ one has
\begin{equation}
\left(\frac{n}{m}\right)_{xx}-P_{xx}(0)=0.
\label{eq:Normal}
\end{equation}
We show below that the appropriate form for NSR-1 is
\begin{eqnarray}\label{eq:nmeq}
\left(\frac{n}{m}\right)_{\alpha\beta} &=& 2\sum_{K}\frac{\partial^{2}\xi_{k}}{\partial K_{\alpha}K_{\beta}}G(K)
\end{eqnarray}
($\alpha,\beta=x,y,z$) and that this is consistent with
the absence of a normal state Meissner effect.

Here the current-current correlation function is
\begin{eqnarray}
P_{\mu\nu}(Q)&=&\int_{0}^{\beta}d\tau e^{i\Omega_{l}\tau}\langle j_{\mu}(\tau,\mathbf{q})j_{\nu}(0,-\mathbf{q}) \rangle \\
&=&-2\sum_{K}\Lambda_{\mu}(K,K_{+})G_{0}(K_{+})\lambda_{\nu}(K_{+},K)G_{0}(K), \nonumber
\end{eqnarray}
where $\lambda$ and $\Lambda$ denote the bare and full vertices and they
necessarily satisfy a Ward identity.  Importantly, as shown in 
Appendix~\ref{App:D},
the two contributions cancel each other as a consequence of a Ward
identity.  This necessary cancellation imposes an important
consistency.
We have presumed that the number
equation appears as in Eq.~(\ref{eq:NSR50}) (which we call NSR-1) which
is then consistent with Eq.~(\ref{eq:nmeq}). If, on the otherhand, we
had assumed the number equation as in NSR-2 , the cancellation can be
enforced as well, but only by proper imposition of the corresponding
Ward identity.  This may explain why the results for the superfluid
density in Refs. \cite{Griffingroup1} and
\cite{Griffingroup2} were not precisely the same as those found in Reference
\cite{Strinati6}.  This analysis also serves to help establish
those diagrams which must be used below $T_c$ in order to be assured
that there are no contributions to the Meissner current from
non-condensed pairs. In view of the above arguments and Appendix~\ref{App:D}, 
the diagrammatic choice in
Reference \cite{Strinati6} seems to be validated, although it is
of interest to reformulate these calculations by explicitly
imposing the
Ward Identity.

The collective mode spectrum appropriate to the NSR scheme was
originally discussed by Griffin and collaborators \cite{Griffin2} based
on the pole structure in Eq.~(\ref{eq:Collective}). This calculation
involves a natural extension of the collective mode calculations
performed at the mean field level \cite{Cote,Randeria2}, but here one
uses the fully self consistent $\mu^*$.  In addition there has been work on
the collective modes using NSR-2
which addresses an improved ground state which
includes quantum fluctuations \cite{Randeriaab}. This, thus, goes beyond
the mean field calculations of this earlier work, and quantifies the
changes in the sound velocity.

\subsection{Alternative Schemes}
\label{sec:4d}

In this Review we have confined our attention to
the two schools of BCS-BEC
crossover which represent natural extensions of the seminal \cite{NSR,Leggett} 
NSR and Leggett papers.
There are alternate approaches which have been introduced into the literature.
Most notable among these is a scheme associated with Zwerger \cite{Zwerger}, 
Haussmann \cite{Haussmann} and their collaborators. The original work \cite{Haussmann}
could be viewed as a third alternative T-matrix scheme in which the
pair propagator $\chi(Q)$ appearing in
Eq.~(\ref{eq:14a})
involves two dressed Green's functions.
In the context of work on high temperature superconductors, this
scheme (and a closely related approach known as ``fluctuation exchange"
or FLEX) 
has been addressed by a number of different groups \cite{Micnas95,YY,Tchern}
and there has been some  
controversy \cite{Tremblay,Tchern,Micnas95} about whether pseudogap
effects naturally emerge. This approach has recently been
extended \cite{Zwerger} below $T_c$
in somewhat the same spirit as the NSR-based schemes. 

\section{
Detailed Comparisons}
\label{sec:5}

\subsection{Overview of Salient Qualitative Comparisons}
\label{sec:comp1}

\begin{table*}[!bht]
\begin{center}
\begin{tabular}{|p{1.3in}|p{2.2in}|p{2.2in}|}
\hline
& \parbox[c][7mm][c]{2.2in}{\textbf{NSR Based Scheme for general $T$}} &
\parbox[c][7mm][c]{2.2in}{\textbf{BCS-Leggett Based Scheme for general $T$}} \\
\hline
\parbox[c][8mm][c]{1.3in}{Fermionic Dispersion Below $T_c$} &
\parbox[c][8mm][c]{2.2in}
{
$\Ek^0= \sqrt{(\ek-\mu^{*})^{2} +\Delta_{sc}^2(T)}$, approximate treatment of fermions
}
& \parbox[c][8mm][c]{2.2in}{$\Ek = \sqrt{(\ek-\mu)^2  + \Delta_{sc}^2(T)
+ \Delta_{pg}^2(T)}$} \\
\hline
\parbox[c][8mm][c]{1.3in}{Bosonic Dispersion Below $T_c$} &
\parbox[c][8mm][c]{2.2in}
{$\Omega_q = c(T)q$}
& \parbox[c][8mm][c]{2.2in}{$\Omega_q^0 = q^2/2M^*$, approximate treatment of bosons} \\
\hline
\parbox[c][8mm][c]{1.3in}{Order of Transition at $T_c$} &
\parbox[c][8mm][c]{2.2in}{First order} &
\parbox[c][8mm][c]{2.2in}{Second order} \\
\hline
\parbox[c][8mm][c]{1.3in}{Density Profiles at Unitarity} &
\parbox[c][8mm][c]{2.2in}{Features indicating condensate edge} &
\parbox[c][8mm][c]{2.2in}{Smooth, quasi-Thomas-Fermi} \\
\hline
\parbox[c][8mm][c]{1.3in}{Superfluid Density}&
\parbox[c][8mm][c]{2.2in}{Multi-valued or Discontinuous at $T_c$}&
\parbox[c][8mm][c]{2.2in}{Smooth and Monotonic at all T}\\
\hline
\parbox[c][8mm][c]{1.3in}{Calculations of Critical Velocity at $T=0$}&
\parbox[c][8mm][c]{2.2in}{}&
\parbox[c][8mm][c]{2.2in}{Either From Vortex or Josephson Effect Calculations}\\
\hline
\parbox[c][8mm][c]{1.3in}{T=0 Superfluid fraction at Unitarity}&
\parbox[c][8mm][c]{2.2in}{100 \%}&
\parbox[c][8mm][c]{2.2in}{100 \%}\\
\hline
\parbox[c][8mm][c]{1.3in}{Order parameter collective modes}&
\parbox[c][8mm][c]{2.2in}{$\omega = c(T) q$
}&
\parbox[c][8mm][c]{2.2in}{$\omega = c'(T) q$
}\\
\hline
\parbox[c][12mm][c]{1.3in}{Major Advantage of Ground State}&
\parbox[c][12mm][c]{2.2in}{Captures physics of Bogoliubov theory, esp. good for BEC}&
\parbox[c][12mm][c]{2.2in}{Allows spatial dependence.
\newline via Bogoliubov deGennes theory} \\
\hline
\end{tabular}
\caption{\label{tab:I}Comparison of Conceptual Issues in the Two Different Theoretical
Schools}
\end{center}
\end{table*}

Because this paper is principally aimed at addressing theoretical issues,
we do not review the vast number of theory-experiment comparisons now
in the literature. These are based on radio frequency spectroscopy,
thermodynamics, collective modes
and other techniques. Rather, here we address
some of the major ``milestone" issues which are often used to assess
the general quality of a given BCS-BEC crossover theory.
We begin with
Table~\ref{tab:I} which presents an overview of the two
theoretical schools as summarized in 
Sections \ref{sec:3}
and
\ref{sec:4}. 
The first two lines characterize the behavior of the fermionic
and bosonic dispersion as they appear in the respective ``gap equations"
of the two schools. 
As is consistent
with the hierarchy of ground state wavefunctions in
Section \ref{sec:2a},
one can infer that the NSR-based scheme
approximates the fermionic contribution and
focuses more directly on the bosonic contribution; it thereby arrives
at a linear dispersion for the pairs. By contrast the BCS-Leggett
school approximates the bosonic contribution 
and focuses more directly on the fermionic
dispersion, thereby incorporating pseudogap effects into
$\Ek$. 
The order of the transition at $T_c$ is second order in the
BCS-Leggett scheme and first order 
\cite{firstordertransitionpapers}
in NSR-based approaches.
The finite $T$ density profiles 
in a trapped gas will reflect this behavior and be rather smooth 
and featureless in the BCS-Leggett scheme \cite{JS5}
while there will be derivative discontinuities
and non-monotonic features \cite{Strinati4} which reflect the condensate edge
in the NSR based scheme.
Similarly the first or second order of the transition will
also show up in the superfluid density within the
BCS-Leggett \cite{JS5,Kosztin1,Chen2} which displays smooth monotonic
behavior or NSR based 
\cite{Strinati6,Griffingroup1,Griffingroup2} scheme which 
shows multi-valed or discontinuous features at $T_c$.
We point out that these spurious first order effects are also seen
in the Bogoliubov theory for true bosons, as discussed in
Appendix \ref{App:C}. 

Calculations of the critical velocity have been
addressed within the BCS-Leggett school
using Bogoliubov deGennes theory
\cite{Hovortex} and from Josephson junction studies
\cite{StrinatiJosephson}. 
Here an experimental comparison can also be made and the
agreement \cite{Ketterle_Criticalcurrent} is reasonable.
Table~\ref{tab:I} shows that in both schools the superfluid
fraction in the ground state is 100 \% in both schools. In
the NSR 
\cite{Cote,Randeria2} and BCS-Leggett \cite{Kosztin2}
schemes the dispersion of the order parameter displays the expected
linear behavior at long wavelengths.
Finally we address the strengths of both ground states by noting
that the NSR-based scheme captures the physics of Bogoliubov theory
and should, thus be the quantitatively better ground state, particularly in the
BEC limit. By contrast the BCS-Leggett scheme is the more flexible
and allows a spatial dependence to be readily incorporated in the form of
Bogoliubov deGennes theory. Moreover, within the BEC, this BdG theory leads
to a Gross Pitaevski picture of the ground state, which allows one
to exploit a
well established body of literature on true Bose systems.

\subsection{Comparison of Superfluid Transition Temperatures}
\label{sec:comp2}

\begin{figure}
\includegraphics[width=2.8in,clip]
{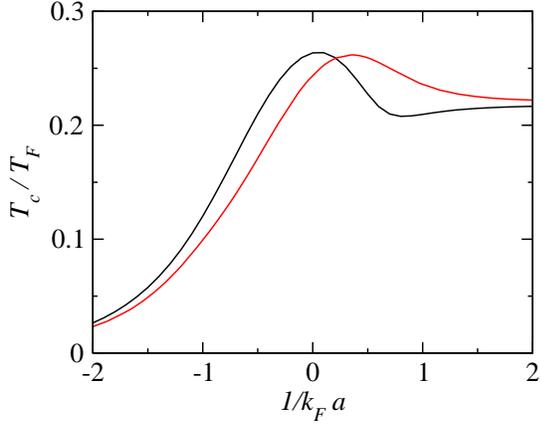}
\caption{ Comparison of $T_c/T_F$ as a function of
inverse scattering length $1/k_Fa$
in a homogeneous
system within the BCS-Leggett scheme
\cite{ourreview,Varenna} (black curve) and the Nozieres Schmitt-Rink
\cite{Strinati4} (red curve) scheme.
The former has a maximum closer to unitarity and a dip close
to the point where $\mu$ changes sign.}
\label{fig:1}
\end{figure}

\begin{figure}
\includegraphics[width=2.8in,clip]
{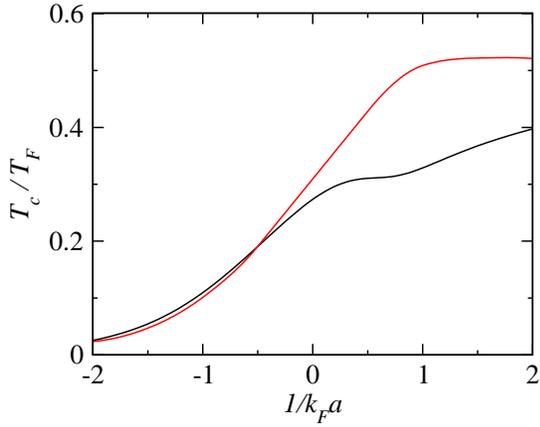}
\caption{ Comparison of $T_c/T_F$ in the trapped case
for the two schools, using the local density
approximation. As in the previous figure, the red
curve is for the NSR scheme \cite{Strinati4} and the black
curve for the BCS-Leggett approach \cite{ourreview,Varenna}.}
\label{fig:2}
\end{figure}

Figures~\ref{fig:1} and \ref{fig:2} present comparisons of the superfluid transition 
temperatures in the two schemes for the homogeneous situation
and in a trap. The black lines correspond to the
BCS-Leggett scheme \cite{ourreview,Varenna} and the red lines are for
the NSR approach as obtained in Reference \cite{Strinati4}.
For the homogeneous case, it can be seen that there are only small
quantitative differences, while in the trapped situation
the BCS-Leggett scheme leads to considerable lower $T_c$ values
slightly above unitarity.
The root of the difference in the two 
calculational schemes lies physically in the fact that the BCS-Leggett
scheme computes the transition temperature in the presence of a finite
(pseudo)gap at $T_c$. In the NSR based scheme, these pair fluctuation
effects do not appear as a pseudogap in the expression for
$T_c$, but rather enter through corrections to the fermionic
chemical potential $\mu^*$.

Section \ref{sec:3a}
presented simple arguments which show
that the ideal gas asymptote
for $T_c$ is approached from below in the BCS-Leggett scheme,
while it evidently is approached from above in the scheme of
Nozieres and Schmitt-Rink. Both of these are mean field approaches 
and the behavior should not be compared with 
expectations \cite{BaymTc} based on a critical fluctuation
description of true Bose systems which clearly include
other physical mechanisms. Indeed, the fact that at $T_c$
there is a discontinuity in the NSR-based schemes suggests that this
approach should be more suitable at $T \approx 0$, away from $T_c$.

Because of the different approaches to the ideal gas asymptote, in
a trap one sees from Figure~\ref{fig:2} that the differences between
the two transition temperatures are
more marked. The ideal gas asymptote is quickly reached in the NSR 
scheme very close to the point where $1/k_F a  \approx 1$. 
In the BCS-Leggett scheme there is an extended regime at and on
the BEC side of
unitarity where $T_c$ 
is rather constant, and the asymptote is only reached for $1/k_Fa$ considerably
larger than its counterpart in the alternate school.

\subsection{Comparison Of Density Profiles}
\label{sec:comp3}

\begin{figure}
\includegraphics[width=2.8in,clip]
{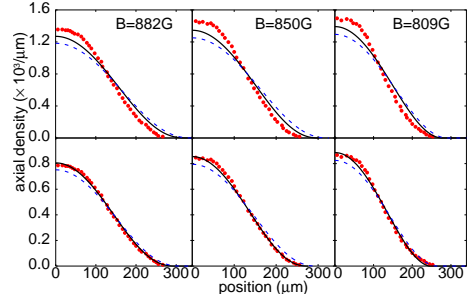}
\caption{Nozieres Schmitt-Rink based results for ground
state profiles. Shown is a comparison between experimental and theoretical
axial density profiles from Ref. \cite{Strinati5}. 
Experimental data from Ref. \cite{Grimm2}
(dots) are shown for three different values of the magentic field B
tuning. Theoretical results for NSR theory \cite{Strinati5} at T=0
(solid lines) and for BCS-Leggett theory (dashed lines) are shown for the
same corresponding parameters. The upper (lower) panels refer to the
estimated number of atoms $N= 4 \times 10^5$ ($N=2.3 \times 10^5$).
}
\label{fig:3}
\end{figure}

Figure \ref{fig:3} presents a plot from Ref. \cite{Strinati5}
of the axial density profiles in
the BCS-Leggett ground state (dashed lines) as compared
with the NSR-derived ground state (black lines) and the
data points (shown in red) for $^6$Li.  
In axial profiles two of the three dimensions of the
theoretical trap profiles were integrated out to obtain a
one-dimensional representation of the density distribution along the
transverse direction: $\bar {n} (x) \equiv \int dydz \, n(r)$.
Three different values of the magnetic field near unitarity
are shown, and
the upper and lower panels correspond to slight changes in the
number of atoms, $N$ which are assumed in the theoretical calculations.
The figure shows that the agreement between theory and experiment
is better for the smaller value of N. 
While the difference in the profiles associated with the two ground
states is not particularly dramatic, it should be stressed that
this difference is reflected in rather large changes in the coefficient
$\beta$ discussed below.
Overall the quantitative agreement between theory and experiment is seen to be
better for the NSR-based ground state.

\begin{figure}
\includegraphics[width=2.8in,clip]{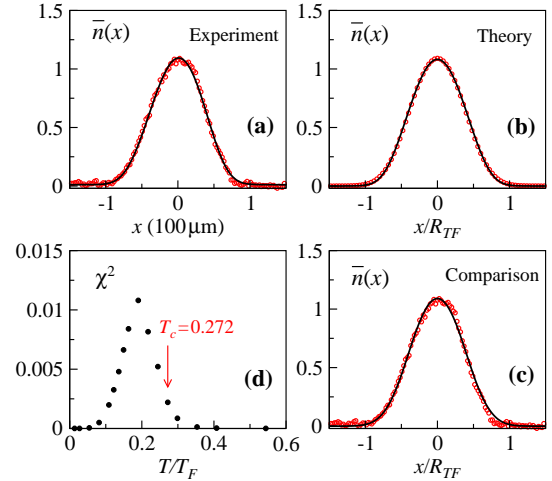}
\caption{(color online) BCS-Leggett results for
temperature dependent profiles of (a) experimental
one-dimensional spatial profiles (circles) and TF fit (line) from
Ref.~\cite{Kinast}, (b) TF fits (line) to theory at $T \approx
0.7T_c \approx 0.19T_F$ (circles) and (c) overlay of experimental
(circles) and theoretical (line) profiles, as well as (d) relative rms
deviations ($\chi^2$) associated with these fits to theory at
unitarity.  The circles in (b) are shown as the line in (c).  The
profiles have been normalized so that $N=\int \bar{n}(x) dx = 1$, and
we set $R_{TF} = 100~\mu$m in order to overlay the two curves.
$\chi^2$ reaches a maximum around $T=0.19T_F$.}
\label{fig:4}
\end{figure}

In Figure~\ref{fig:4} are shown density profiles at finite
temperatures for the BCS-Leggett case, from
Reference \cite{JS5}.
The experimental data and theory correspond to roughly
$T/T_F = 0.19$.
These profiles are estimated to be within the
superfluid phase ($T_c \approx 0.3T_F$ at unitarity).  This figure
presents Thomas-Fermi fits \cite{Kinast} to the experimental
(\ref{fig:4}a) and theoretical (\ref{fig:4}b) profiles as well as
their comparison (\ref{fig:4}c), for a chosen $R_{TF} = 100~\mu m$,
which makes it possible to overlay the experimental data (circles) and
theoretical curve (line).  Finally Fig.~\ref{fig:4}d indicates the
relative $\chi^2$ or root-mean-square (rms) deviations for these TF
fits to theory.  This figure was made in collaboration with the
authors of Ref.~\cite{Kinast}.
To probe the deviations from a TF functional form, 
in Fig.~\ref{fig:4}d,  
the
(relative) rms deviation, or $\chi^2$, from the TF fits as a function
of $T$ is plotted.  
$\chi^2$ increases rapidly below $T_c$ and reaches a
maximum around $0.7T_c$.  
Quite good agreement between theory and experiment is observed here
in the finite $T$ profiles.

\subsection{Addressing $\beta$ Effects}
\label{sec:5a}

One of the most widely used milestones for assessing crossover theories is the
numerical value obtained for the coefficient $\beta$.
At unitarity the chemical potential must scale with the Fermi energy
with a coefficient of proportionality
\begin{equation}
 \mu = ( 1+ \beta ) E_F .
\label{eq:77}
\end{equation}
In the BCS-Leggett ground state $\beta \approx -0.41$.  By contrast
experimental data \cite{Grimm2} suggest an answer which is closer to
Monte Carlo calculations \cite{carlson3} $\beta = -0.56$.  Calculations
\cite{Strinati5} based on NSR-1 yield $\beta = -.545$, which is in quite
good agreement with experiment.  In the NSR-2 scheme
(based on the thermodynamical potential with the non-variational
contribution included), the same good agreement with experiment (and
Monte Carlo) was presented
\cite{Drummond2,Drummond3,Randeriaab}.

The BCS-Leggett- based scheme, as it has
been implemented here, can be seen to ignore Hartree effects, just as is
consistent with the ground state wavefunction.  This is a shortcoming
of the scheme and in the context of the formalism presented
here,  one can trace it to
Eq.~(\ref{eq:sigma3}). This approximation,
in effect, includes only pairing correlations in the self energy. Correlations
that are not associated with pairing, such as Hartree effects have been
omitted.
With the full T-matrix formalism as outlined in
Section \ref{sec:3f},
it should be clear that
this assumption can be avoided and is not fundamental to the physical
picture presented here.  However,
dropping this simplification does lead to considerable numerical complexity.
We note that the NSR-based theories both above and below $T_c$ include these
non-pairing correlations
in a fairly automatic way.
In some limited contexts, they have also been included in the BCS-Leggett based
theory \cite{Janko,Maly1,Maly2}.

One can think of these omitted correlations as entering
via
Eq.~(\ref{eq:14a})
when the pair susceptibility is assumed \cite{BaymBruun}
to include only two bare Green's functions.
Most recently, it has been
shown that these ``$G_0G_0$" correlations are responsible for some important
physical observations
in the context of gases which are so
strongly polarized that superfluidity is driven away
\cite{KetterleRF}.
A bound state associated with the minority spins is found to occur \cite{Lobo,Chevy2}
in these highly imbalanced gases, which is responsible \cite{Punk,Stoof3}
for anomalies in the RF
spectra \cite{KetterleRF}.

In summary, it is possible to estimate the size of these
Hartree corrections, if one goes beyond Eq.~(\ref{eq:sigma3}) and
includes the effects deriving from ``$G_0G_0$" correlations, noted above.
This will have to be done in future calculations for better quantitative
comparisons of various properties, including $\beta$.

\begin{table*}[!bht]
\begin{center}
\begin{tabular}{|p{1.3in}|p{1.3in}|p{1.3in}|p{1.3in}|p{1.3in}|}
\hline
& \parbox[c][7mm][c]{1.3in}{\textbf{NSR -1}}&
\parbox[c][7mm][c]{1.3in}{\textbf{NSR-2}}&
\parbox[c][7mm][c]{1.3in}{\textbf{BCS-Leggett}}&
\parbox[c][7mm][c]{1.3in}{\textbf{``Answer''}} \\
\hline
\parbox[c][8mm][c]{1.3in}{scatt. length ratio: $a_B/a$}&
\parbox[c][8mm][c]{1.3in}{2.0 } &
\parbox[c][8mm][c]{1.3in}{$\approx$ 0.55 } &
\parbox[c][8mm][c]{1.3in}{2.0} &
\parbox[c][8mm][c]{1.3in}{0.6 (exact calc.)  }\\
\hline
\parbox[c][8mm][c]{1.3in}{$\beta = \frac{\mu}{E_F} -1$ }&
\parbox[c][8mm][c]{1.3in}{-0.545 } &
\parbox[c][8mm][c]{1.3in}{-0.59 } &
\parbox[c][8mm][c]{1.3in}{-0.41} &
\parbox[c][8mm][c]{1.3in}{-0.55, (from experiment)}\\
\hline
\parbox[c][8mm][c]{1.3in}{$N_c$ at T=0, $1/k_Fa = 2$ }&
\parbox[c][8mm][c]{1.3in}{} &
\parbox[c][8mm][c]{1.3in}{0.84 } &
\parbox[c][8mm][c]{1.3in}{0.99} &
\parbox[c][8mm][c]{1.3in}{0.96 (Monte Carlo)} \\
\hline
\parbox[c][8mm][c]{1.3in}{$N_c$ at T=0, $1/k_Fa = \infty$ }&
\parbox[c][8mm][c]{1.3in}{} &
\parbox[c][8mm][c]{1.3in}{0.48 } &
\parbox[c][8mm][c]{1.3in}{0.69} &
\parbox[c][8mm][c]{1.3in}{0.58 (Monte Carlo)} \\
\hline
\end{tabular}
\caption{\label{tab:II}Quantitative Comparisons among the different schools. References
for each number are listed in the first row from left to right:
Ref.~\cite{PS05}, Ref.~\cite{Drummond3}, Ref.~\cite{Stringari},
and Ref.~\cite{Petrov}.  For the second row from left to right
the references are: Ref.~\cite{Strinati5}, Ref.~\cite{Drummond3},
Ref.~\cite{Stringari} and Ref.~\cite{Grimm2}. For the third row
from left to right the references are Ref.~\cite{Griffingroup2},
Ref.~\cite{MCpair} and Ref.~\cite{MCpair}.
Finally in the last row from left to right the references are
Ref.~\cite{Griffingroup2},
Ref.~\cite{MCpair} and Ref.~\cite{MCpair}.}
\end{center}
\end{table*}

\subsection {Condensate Fraction and ``Quantum Depletion''}
\label{sec:5b}

It is interesting to contemplate the concept of ``quantum depletion'' in
a fermionic system, particularly as one approaches the BEC.  It is generally
believed that the BCS-Leggett theory is to be distinguished from
that based on NSR (which is closer to Bogoliubov theory) because
of the neglect of quantum depletion. However, because of
the presence of unpaired fermions, 
the condensate fraction will automatically
show less than 100\%
condensation, except in the deepest BEC. Similarly, in the BCS regime this
condensate fraction is vanishingly small. Establishing
the degree to which 
``quantum depletion'' is present in a Fermi gas is, thus, a subtle
issue.

We begin with the BCS-Leggett ground state.  Following earlier work
\cite{Yang}, at $T=0$, the pair wavefunction is defined as $F_{\bf
  k}\equiv\langle N-2|c_{-{\bf k}\downarrow}c_{{\bf
    k}\uparrow}|N\rangle$, where $c_{{\bf k}\sigma}$ is the fermion
annihilation operator for $\sigma=\uparrow, \downarrow$.  It can be
shown that in this ground state we have $F_{\bf k}=u_{\bf k} v_{\bf k}$, where
 the coefficients are $u^2_{\bf k},v^2_{\bf k}=[1\pm(\ek-\mu)/E_{\bf k}]/2$ and $E_{\bf k}=\sqrt{(\ek-\mu)^2+\Delta^2}$. The condensate
fraction at $T=0$ is
\begin{equation}\label{eq:Yangformula}
N_c=\sum_{\bf k}|F_{\bf k}|^2=\int d{\bf r}|F({\bf r})|^2 .
\end{equation}
Here $F({\bf r})=\sum_{\bf k}F_{\bf k}\exp(i{\bf k}\cdot{\bf r})$. This
pair density reflects off-diagonal-long-range-order.  There have been a
number of numerical calculations of this quantity over the entire
BCS-BEC crossover \cite{MCpair} and the agreement with direct Monte
Carlo schemes \cite{MCpair} is not
unreasonable, as will be summarized below.

It is natural to try to extend this picture to finite temperature,
taking the quantity $F_{\bf k}=T\sum_{\omega_n}F(i\omega_n,{\bf k})
=u_{\bf k}v_{\bf k}[1-2f(E_{\bf k})]$ as a measure of the pair density. We stress that
this is not related to off-diagonal long range order, but rather
contains the contributions from condensed and non-condensed pairs,
through the decoupling of $\Delta^2$ into $\Delta_{sc}^2$ and
$\Delta_{pg}^2$. One has, thus,
\begin{equation}
n_{pair}=\Delta^2[1-2f(E_{\bf k})]^2/4E_{\bf k}^2, ~~~ T \neq 0.
\end{equation}


To emphasize that there is no unique representation of the pair fraction
away from the BEC limit, we note that Eq.~(\ref{eq:25}) provides
another natural decomposition
We can rewrite this equation representing
the total density of fermions
$n$
in the form
\begin{equation}
n= 2 Z \Delta^2
+2\sum_{\mathbf{k}}f(\ek-\mu)
\label{eq:25a}
\end{equation}
or equivalently
\begin{equation}
n=2Z\Delta_{sc}^2+2Z\Delta_{pg}^2+2\sum_{\mathbf{k}}f(\ek-\mu),
\label{eq:decomp}
\end{equation}
from which
\begin{equation}
n_{pair}=2Z\Delta_{sc}^2+2Z\Delta_{pg}^2,
\end{equation}
can be obtained.  There are three terms on the right hand side of
Eq.~(\ref{eq:decomp}).
The second term corresponds to the density of fermions in the
non-condensed pairs.  The first term may be identified 
as
$N_c=2Z\Delta_{sc}^2$, representing an alternative way of
quantifying the density of fermions in the condensate, and the third
term may be identified as the density of remaining (unpaired) fermions,
$n_f=2\sum_{\mathbf{k}}f(\ek-\mu)$.  This decomposition is of interest,
in part because it relates more directly to the decomposition of pairing
contributions and free fermions introduced in the original NSR paper.

Recent calculations of $N_c$ \cite{Griffingroup1,Griffingroup2} have
also been presented using the NSR-2 approach, where the fraction is
found to be somewhat smaller than in the BCS-Leggett state. Importantly,
the difference between these two results for $N_c$ is viewed as a
possible way to represent quantum depletion, which is naturally larger
in NSR based theories as compared to the BCS-Leggett counterpart.

\subsection{Effects of First Order Transitions}
\label{sec:5c}

Essentially all NSR-based theories, as well as some which claim higher levels
of consistency, report first order transitions
\cite{firstordertransitionpapers}.  These effects presumably originate
in the same way as their counterparts in true Bose systems treated at
the Bogoliubov \cite{Griffin4} or Popov level.  
We outline the origin of these first order effects in
Appendix \ref{App:C}.
They lead to derivative
discontinuities in the density profiles at the condensate edge
\cite{StringariGiorgini} and are thus, not as problematic in the case of
a trapped gases as compared to a homogeneous system.  This is
particularly the case in the BEC where bimodality is present and one
would expect signatures of the condensate edge.

Despite this theoretical framework,
experiments show a behavior which is far from first order.
One of the most striking features about the unitary gases is
that there is so little indication of the phase
transition and thus no evidence for first order behavior.
This is seen by noting 
the historical difficulties encountered in establishing whether a particular experiment is
performed in the superfluid or normal phase. In the absence of
population imbalance, the unitary gas profiles are featureless
\cite{ThermoScience} with no clear bi-modality or other indications of a
condensate edge.  Similarly, RF spectroscopic studies of the pairing gap
show a smooth behavior \cite{Grimm4} from high $T$ to temperatures well
below $T_c$.  As a consequence, $T_c$ is difficult to identify, although
important thermodynamical measurements, have indeed, indicated a phase
transition \cite{ThermoScience}.

These theoretically
generated first order effects become even more difficult to reconcile with
the fact that in BCS-
BEC crossover, the pseudogap, which
appears well above $T_c$ leads to an even smoother transition
than in strict BCS theory (which also is of second order). 
Thus a first order
transition in systems undergoing BCS-BEC crossover can be viewed as
somewhat problematic, except, perhaps if attention is restricted to a
narrow temperature range.  This points to an advantage of the
BCS-Leggett based approach where the density profiles are rather
featureless and well fit to a Thomas Fermi form.  A related advantage is
that without first order transitions one can arrive at a theoretical
basis \cite{Carr,ChenThermo} for adiabatic sweep thermometry. This is an
experimental technique \cite{Jin_us,Ketterle3} which has been rather
widely discussed.  Using the theoretically determined entropy, it is
possible to arrive at reasonable estimates of a final
temperature, based on an experimentally known initial temperature
connected by an adiabatic sweep.

\subsection{Quantitative Comparisons}

Quantitatively, the NSR-based approaches appear to have some advantage,
although there are variations depending on how the number equation
is implemented (either via NSR-1 or NSR-2).
Table~\ref{tab:II} summarizes specific key numbers which have been used to assess
the different schools.
Listed in the first row is the scattering length ratio associated
with the inter-boson and inter-fermion interactions, while the second
row compares the quantity $\beta$
defined in Eq.~(\ref{eq:77}) which is associated with
the unitary limit. Finally, the third and fourth rows address
the value of the condensate fraction in the ground state in the near-BEC
and very deep BEC. This is relevant to quantifying the degree of
quantum depletion.
We have previously addressed our concerns about NSR-2 which,
through 
Eq.~(\ref{eq:75b}), builds on inconsistencies associated
with the fact that the gap equation does not satisfy the variational
condition.
Rather we argue in favor of
the approach we call NSR-1 which uses Eq.~(\ref{eq:NSR50}).
While there seems to be considerable interest in the community in
comparing numbers such as those shown in Table~\ref{tab:II}, because of just these
concerns about more fundamental issues, we are of the
opinion that it may be premature to give too much weight to
the numerical
comparisons shown in Table~\ref{tab:II}. Instead we attach greater
importance to Table~\ref{tab:I} for assessing a given theory and for indicating new
directions for improvement.

\section{Summary}
\label{sec:6}

One of the major goals of this review has been to clarify the genesis
of a large number of contributions to the theoretical
literature by associating them clearly 
with one or the other theoretical approaches to BCS-BEC crossover.
We stress 
that these two theoretical schemes are different in the ground state and
in their thermal properties. One should, thus, avoid the tendency to
present results from the Leggett BCS ground state and simultaneously
use the Nozieres Schmitt-Rink calculations for treating
$ T \neq 0$ aspects of BCS-BEC crossover.

It was also our goal to summarize the major strengths and weaknesses
within these two schools. While it clearly includes bosonic
degrees of freedom, 
there is a concern about the Leggett-BCS theory
which concern derives from the fact that
this approach does not yield a Bogoliubov-like or sound
mode dispersion for the \textit{non-condensed pair
excitations}. 
Formally, this is a consequence of the associated
T-matrix description of the
$q \neq 0$ pairs, which drops higher order terms and which
are needed to couple the condensate
and pair excitations.
Although it has not been seen as yet \cite{FOOTNOTE},
this sound-like excitation spectrum could show up in
future experiments on unitary gases, particularly through power law
dependences in thermodynamics.
On the positive side, calculations in this BCS-Leggett phase are
very tractable; one can readily handle inhomogeneities such as vortices
(through the Bogoliubov deGennes approach); one can introduce trap effects, 
as well as population imbalance and address all temperatures $T$.

For the NSR based schemes a comparably major problem is that there is no
satisfactory mean field theory for the weakly interacting Bose
gas which works at all temperatures. The NSR based school is based on this
Bose gas mean field starting point and this introduces
unphysical first order transitions which, at least around $T_c$, 
will interrupt the smooth crossover from BCS to BEC and
limit the applicability of theory 
to specified ranges of temperature.
On the positive side, it is believed that this scheme, which works
best at low temperatures, will produce the better ground state and
allow more quantitative comparison with experiment at $T \approx 0$.

With these weaknesses identified, each of these schools has a large
agenda in hand for future research. 
In the short term the BCS-Leggett scheme should be readily extended to include
additional non-pairing contributions to the self energy (such as
Hartree effects) which will
make it more favorable for quantitative comparisons. Similarly, in the
short term, the NSR-based approach should be extended to implement
the inclusion of
Popov like correlations, and thereby include corrections to
the BCS gap equation (which treats the fermions
as non-interacting). In the longer term one would hope that
NSR scheme (which approaches the
crossover from the BEC end and oversimplifies the fermionic
dispersion) and the BCS-Leggett scheme (which approaches
the crossover form the BCS end and oversimplifies
the bosonic dispersion), will ultimately be unified. It is
also to be expected that experiments will guide the way.

\appendix

\section{Details of the T-matrix calculations}
\label{App:A}
The T-matrix is $t_{pg}^{-1}(Q)=[U^{-1}+\chi(Q)]$. Near $T_c$ can it be
expanded near $Q=0$ in the form
\begin{equation}
t_{pg}(\mathbf{q},\Omega)=\frac{1}{Z(\Omega-q^2/2M^*)+\mu_{pair}+i\Gamma_Q}
\end{equation}
after analytic continuation ($i\Omega_l\rightarrow\Omega+i0^+$). The pair chemical potential $\mu_{pair}$ vanishes below $T_c$. The relaxation term $i\Gamma_Q$ is neglected near $T_c$ in most applications.
 Firstly we calculate $Z$.
 \begin{eqnarray}
 Z=\frac{\partial t_{pg}^{-1}}{\partial\Omega}\Big|_{\Omega=0,q=0}=\frac{1}{2\Delta^2}\left[n-2\sum_{\mathbf{k}}f(\ek-\mu)\right].
 \end{eqnarray}
 The effective mass of pairs, $M^*$, is given by
 \begin{eqnarray}
 \frac{1}{2M^*}&=&\frac{1}{6Z}\frac{\partial^{2}t^{-1}_{pg}(\mathbf{q},0)}{\partial \mathbf{q}^{2}}\Big|_{q=0}.
 \end{eqnarray}
 Here
 \begin{widetext}
 \begin{eqnarray}
 \frac{\partial^{2}t^{-1}_{pg}(\mathbf{q},0)}{\partial \mathbf{q}^{2}}\Big|_{q=0}
 &=&-\frac{1}{2\Delta^{2}}\sum_{\mathbf{k}}\Bigg\{2f(\ek-\mu)\Big[(\nabla_{\bf k}^{2}\ek)+4\Big(\frac{\ek-\mu}{\Delta^{2}}\Big)(\nabla_{\bf k}\ek)^{2} \Big]  -2f(E_{\bf k})\Big[\Big(\frac{\ek-\mu}{E_{\bf k}}\Big)(\nabla_{\bf k}^{2}\ek)+\nonumber \\
& &2\Big\{\frac{(E_{\bf k}^{2}+(\ek-\mu)^{2})}{\Delta^{2}E_{\bf k}}\Big\}(\nabla_{\bf k}\ek)^{2} \Big] +4f^{\prime}(\ek-\mu)(\nabla_{\bf k}\ek)^{2}-\Big(1-\frac{\ek-\mu}{E_{\bf k}}\Big)(\nabla_{\bf k}^{2}\ek)+ \nonumber \\
& &\Big(\frac{2E_{\bf k}}{\Delta^{2}}\Big)\Big(1-\frac{\ek-\mu}{E_{\bf k}}\Big)^{2}(\nabla_{\mathbf{\bf k}}\ek)^{2}\Bigg\}.
 \end{eqnarray}
 \end{widetext}

\section{Details on Eqs. (\ref{eq:60}) and (\ref{eq:62})}
\label{App:B}
We review how Eqs. (\ref{eq:60}) and (\ref{eq:62}) are derived following \cite{PS05}. The Dyson's equation, Eq.(32) in Ref.\cite{PS05}, is
\begin{widetext}
\begin{eqnarray}
\hat{G}_{11}(K)&=&-\hat{G}_{22}(-K)=\hat{G}^{o}_{11}(K)+\hat{G}^{o}_{11}(K)[\Sigma_{11}(K)\hat{G}_{11}(K)+\Sigma_{12}(K)\hat{G}_{21}(K)], \nonumber \\
\hat{G}_{12}(K)&=&\hat{G}_{12}(K)=\hat{G}^{o}_{11}(K)[\Sigma_{11}(K)\hat{G}_{12}(K)+\Sigma_{12}(K)\hat{G}_{22}(K)].
\end{eqnarray}
\end{widetext}

Following the approximation shown in Eq.~(\ref{eq:61}), the self-energy becomes
\begin{eqnarray}
\Sigma_{11}(K)&\approx&\frac{\bar{\Delta}_{pg}^{2}}{i\omega_{n}+\xi_{\bf k}},\mbox{ }\Sigma_{22}(K)\approx\frac{\bar{\Delta}_{pg}^{2}}{i\omega_{n}-\xi_{\bf k}}.
\end{eqnarray}
In the BEC limit, $\Delta_{sc}/|\mu^*|\ll 1$ and we assume this also holds for $\bar{\Delta}_{pg}/|\mu^*|$. Then $\hat{G}_{11}(K)$
\begin{eqnarray}
&=&\frac{i\omega_{n}+\xi_{\bf k}-\frac{\bar{\Delta}_{pg}^{2}}{i\omega_{n}-\xi_{\bf k}}}{\left(i\omega_{n}-\xi_{\bf k}-\frac{\bar{\Delta}_{pg}^{2}}{i\omega_{n}+\xi_{\bf k}} \right)\left(i\omega_{n}+\xi_{\bf k}-\frac{\bar{\Delta}_{pg}^{2}}{i\omega_{n}-\xi_{\bf k}} \right)-\Delta_{sc}^{2}} \nonumber \\
&=&-\frac{(i\omega_{n}+\xi_{\bf k})(\omega_{n}^{2}+\xi_{\bf k}^2+\bar{\Delta}_{pg}^{2})}{(\omega_{n}^{2}+\xi_{\bf k}^2+\bar{\Delta}_{pg}^{2})^{2}+\Delta_{sc}(\omega_{n}^{2}+\xi_{\bf k}^{2})} \nonumber \\
&=&-\frac{i\omega_{n}+\xi_{\bf k}}{\omega_{n}^{2}+\xi_{\bf k}^2+\bar{\Delta}_{pg}^{2}+\Delta_{sc}^{2}\frac{\omega_{n}^{2}+\xi_{\bf k}^2}{\omega_{n}^{2}+\xi_{\bf k}^2+\bar{\Delta}_{pg}^{2}}} \nonumber \\
&\approx&-\frac{i\omega_{n}+\xi_{\bf k}}{\omega_{n}^{2}+\xi_{\bf k}^2+\bar{\Delta}_{pg}^{2}+\Delta_{sc}^{2}}.
\end{eqnarray}
Here we made the approximation, which is valid in the BEC limit,
\begin{eqnarray}
\frac{\omega_{n}^{2}+\xi_{\bf k}^2}{\omega_{n}^{2}+\xi_{\bf k}^2+\bar{\Delta}_{pg}^{2}}=1-\frac{\bar{\Delta}_{pg}^{2}}{\omega_{n}^{2}+\xi_{\bf k}^2+\bar{\Delta}_{pg}^{2}}\approx 1.
\end{eqnarray}
The off-diagonal fermion Green's function $\hat{G}_{12}(K)$
\begin{eqnarray}
&=&-\frac{\Delta_{sc}}{\left(i\omega_{n}-\xi_{\bf k}-\frac{\bar{\Delta}_{pg}^{2}}{i\omega_{n}+\xi_{\bf k}} \right)\left(i\omega_{n}+\xi_{\bf k}-\frac{\bar{\Delta}_{pg}^{2}}{i\omega_{n}-\xi_{\bf k}} \right)-\Delta_{sc}^{2}} \nonumber \\
&=&\frac{\Delta_{sc}(\omega_{n}^{2}+\xi_{\bf k}^2)}{(\omega_{n}^{2}+\xi_{\bf k}^{2}+\bar{\Delta}_{pg}^{2})^{2}+\Delta_{sc}^{2}(\omega_{n}^{2}+\xi_{\bf k}^2)} \nonumber \\
&=&\frac{\Delta_{sc}}{\omega_{n}^{2}+\xi_{\bf k}^{2}+2\bar{\Delta}_{pg}^{2}+\Delta_{sc}^{2}+\frac{\bar{\Delta}_{pg}^{4}}{(\omega_{n}^{2}+\xi_{\bf k}^2)}} \nonumber \\
&\approx&\frac{\Delta_{sc}}{\omega_{n}^{2}+\xi_{\bf k}^{2}+2\bar{\Delta}_{pg}^{2}+\Delta_{sc}^{2}}.
\end{eqnarray}
Note that the corresponding gap equation, Eq.~(\ref{eq:Strinatigap}), derived from this expression is different from the BCS gap equation when $E_{\bf k}$ is defined as $\sqrt{\xi_{\bf k}^{2}+\Delta_{sc}^{2}+\bar{\Delta}_{pg}^{2}}$.

\section{First order transitions in boson mean field theories}
\label{App:C}
It is well known \cite{Giorginiboson,AndersenRMP} that mean 
field theories of the weakly interacting Bose gas are associated with
unphysical first order 
transitions. It is, thus, often argued that these theories should only be
applied 
at temperatures much lower than $T_c$. 
Since the same issues arise with BCS-BEC crossover theories of the extended
Nozieres Schmitt-Rink school (based on
the Hartree-Fock-Bogoliubov 
or Popov approximations), it is useful to understand the physical origin of
these first order effects. 

We 
summarize here the central issues which lead to first order transitions:
\begin{enumerate}
\item The BEC transition temperature predicted by mean field theories 
is the same as the BEC temperature of an ideal Bose gas, $T^{0}_{BEC}$.
\item Below $T^{0}_{BEC}$, interaction effects are found to
\textit{suppress thermal excitations}. This suppression arises 
from interaction effects in
the dispersion relation which lead to a systematic increase in
the excitation energy, relative to the non-interacting gas. In addition
there is a change in the phase space weighting factor. In combination,
these two effects
importantly yield a smaller fraction of non-condensed bosons (or a larger 
condensate fraction).
\item As a consequence, if one plots the condensate
fraction obtained from generic mean field 
theory as a function of $T$, one sees that it tends towards $T^{0}_{BEC}$ by
overshooting and then bending back towards 
$T^{0}_{BEC}$
at the highest temperatures below the transition.
This double valued behavior is then associated with a first order transition.
\end{enumerate}

\begin{figure}
\centerline{\includegraphics[clip,width=2.8in]{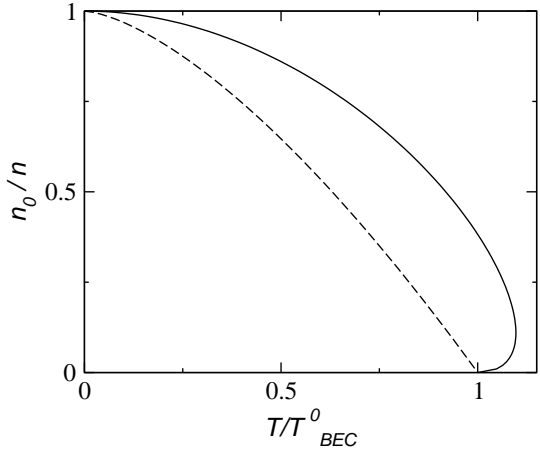}}
\caption{\label{fig:Pop} Condensate fraction $n_0/n$ as a function of
dimensionless temperature $T/T^{0}_{BEC}$. Solid and dashed lines correspond to 
results obtained from 
the Popov approximation and the theory of non-interacting bosons, respectively.}
\end{figure}

Fig.~\ref{fig:Pop} shows
the condensate fraction (solid line)
as a function of temperature as obtained from the
Popov approximation. The bend-over which indicates a first order transition can be
seen clearly. As a comparison, the condensate fraction of a
non-interacting gas of bosons is
also presented (dashed curve). Here one sees a smooth second order phase transition.

We can take these ideas over to the BCS-Leggett approach to BCS-BEC crossover (which
is the only case where a first order transition is not seen). It is rather
straightforward to see the analogies with the Bose gas through the
gap equation:
$\Delta^{2}=\Delta_{sc}^{2}+\Delta_{pg}^{2}$ 
 in conjunction with Eq.~(\ref{eq:18}) (and in some situations also with Eq.~(\ref{eq:24})).
Here, too, following the interacting Bose gas logic we will also end up with
an unphysical first order transition which means that the smooth crossover
at finite $T$ is interrupted for some range of temperatures below and near
$T_c$.

\section{Ward Identity Analysis of the number equation}
\label{App:D}
\begin{figure} 
\centerline{\includegraphics[clip,width=3.in]{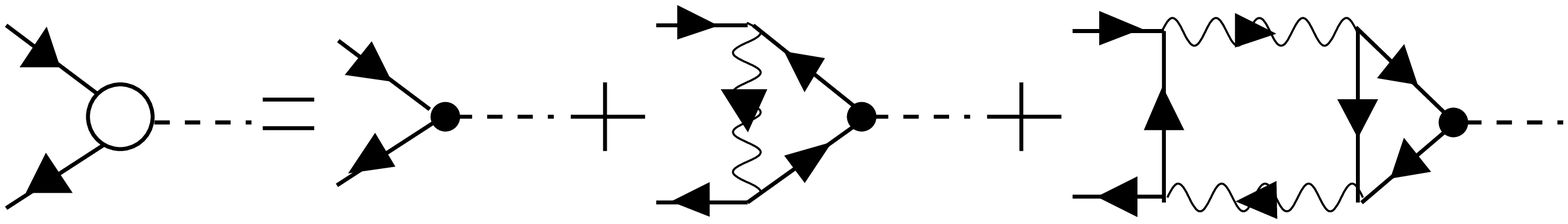}}
\caption{\label{fig:vertex} The full vertex is approximated by these diagrams. The first one on the right hand side is the bare vertex, the second one is the ``MT'' diagram, and the last one is the ``AL'' diagram. Hollow and solid dots denote full and bare vertices. Solid lines and wavy lines correspond to propagator of non-interacting fermions and t-matrix, respectively.}
\end{figure}

The Ward identity is given by
\begin{eqnarray}
Q\cdot\lambda(K,K_{+})&=&G_{0}^{-1}(K)-G_{0}^{-1}(K_{+}), \nonumber \\
Q\cdot\Lambda(K,K_{+})&=&G^{-1}(K)-G^{-1}(K_{+}).
\end{eqnarray}
Here $G_{K}=[G_{0}^{-1}(K)-\Sigma(K)]^{-1}$ is the full fermion Green's
function. The full vertex is obtained by choosing
a set of diagramms consistent with gauge invariance. These
are shown
in Fig.~\ref{fig:vertex} and specified in
the caption.
Now we would like to show that this set of diagrams satisfy the Ward
identity and does not contribute to the Meissner effect if the correct
form of number equation is used.

The expressions of those diagrams in Fig.~\ref{fig:vertex} are
\begin{eqnarray}\label{eq:G0G0eq}
\lambda(K,K_{+})&=&(2\mathbf{k}+\mathbf{q}, 1), \\
MT(K,K_{+}) &=& \sum_{P}t(P)G_{0}(P-K_{+})\times  \nonumber \\
& &\lambda(P-K_{+},P-K)G_{0}(P-K), \nonumber \\
AL(K,K_{+}) &=& -2\sum_{L,P}t(P)t(P_{+})G_{0}(P-K)\times  \nonumber \\
& &G_{0}(P-L)G_{0}(L)\lambda(L,L_{+})G_{0}(L_{+}). \nonumber
\end{eqnarray}
The factor $2$ in the AL diagram comes from the fact that the vertex can
be inserted in one of the two particle propagators in the T-matrix. By
taking inner produc with $Q$ they becomes
\begin{eqnarray}
Q\cdot MT(K,K_{+})&=&\sum_{P}t(P)G_{0}(P-K_{+})[G_{0}^{-1}(P-K_{+})- \nonumber \\
& &G_{0}^{-1}(P-K)]G_{0}(P-K), \nonumber \\
&=& -[\Sigma(K_{+})-\Sigma(K)].
\end{eqnarray}
\begin{widetext}
\begin{eqnarray}
Q\cdot AL(K,K_{+}) &=& -2\sum_{L,P}t(P)t(P_{+})G_{0}(P-K)G_{0}(P-L)G_{0}(L)[G_{0}^{-1}(L)-G_{0}^{-1}(L_{+})]G_{0}(L_{+}), \nonumber \\
&=& -2\sum_{P}t(P)t(P_{+})G_{0}(P-K)[\chi(P_{+})-\chi(P)], \nonumber \\
&=& 2[\Sigma(K_{+})-\Sigma(K)].
\end{eqnarray}
\end{widetext}
In deriving these results, Eqs.(\ref{eq:G0G0eq}) and the identity
$\chi(P_{+})-\chi(P)=t^{-1}(P_{+})-t^{-1}(P)$ are useful. Therefore
$Q\cdot(MT+AL)=\Sigma(K_{+})-\Sigma(K)$. It is straighforward to show
that the approximated vertex $\Lambda=\lambda+MT+AL$ satisfies the Ward
identity. Since these diagrams are normal state diagrams, one can take
the limit $Q\rightarrow 0$ and obtain
\begin{equation}
[MT(K,K)+AL(K,K)]_{\mu}=\frac{\partial\Sigma(K)}{\partial K_{\mu}}.
\end{equation}
This is an important identity in the derivation of the absence of
Meissner effect in the normal state.

The Meissner effect occurs if the static response kernel does not
vanish, which is equivalent to the existence of a superfluid density. To
show that the approximation for the full vertex does not contribute to
the Meissner effect, it suffices to show that in the normal state
\begin{equation}
\left(\frac{n}{m}\right)_{xx}-P_{xx}(0)=0.
\end{equation}
Here the density is calculated as ($\alpha,\beta=x,y,z$)
\begin{eqnarray}\label{eq:nmeqa}
\left(\frac{n}{m}\right)_{\alpha\beta} &=& 2\sum_{K}\frac{\partial^{2}\xi_{k}}{\partial K_{\alpha}K_{\beta}}G(K) \\
&=&-2\sum_{K}\frac{\partial\xi_{k}}{\partial K_{\alpha}}\frac{\partial G(K)}{\partial K_{\beta}} \nonumber \\
&=&-2\sum_{K}\frac{\partial\xi_{k}}{\partial K_{\alpha}}G^{2}(K)\left[\frac{\partial\xi_{k}}{\partial K_{\beta}}+\frac{\partial\Sigma(K)}{\partial K_{\beta}}\right]. \nonumber
\end{eqnarray}
We assume that surface terms can be neglected. The current-current
correlation function at $Q=0$ is
\begin{eqnarray}
  P_{\alpha\beta}(0)&=&-2\sum_{K}G^{2}(K)[\lambda(K,K)+MT(K,K)+ \nonumber \\
  & &AL(K,K)]_{\beta}\lambda(K,K)_{\alpha} \nonumber \\
  &=&-2\sum_{K}G^{2}(K)\frac{\partial\xi_{k}}{\partial K_{\alpha}}\left[\frac{\partial\xi_{k}}{\partial K_{\beta}}+\frac{\partial\Sigma(K)}{\partial K_{\beta}}\right],
\end {eqnarray}
where we used $\lambda(K,K)_{\alpha}=
\frac{\partial\xi_{k}}{\partial K_{\alpha}}$. 

Thus the two contributions cancel each other and there is no Meissner
effect in the normal state.

This work is supported by NSF PHY-0555325 and NSF-MRSEC Grant
No.~DMR-0820054. We thank G.C. Strinati, A. Perali, P. Pieri, R.Hulet,
and D. Wulin for helpful communications.

\vspace*{-1ex} 

\bibliographystyle{apsrev}


\begin{thebibliography}{115}
\bibitem[{FOO()}]{FOOTNOTE}
\bibinfo{note}{Although we note that thermodynamic measurements in trapped
  gases have not been able to detect unambiguously the sound mode signatures in
  entropy since most of the weight comes from free fermions at the trap edge
  \cite{ThermoScience}.}
\end{thebibliography}

\end{document}